\documentclass[11pt]{article}
\usepackage{hyperref}
\usepackage{url}
\usepackage[anythingbreaks=true]{breakurl}
\hypersetup{colorlinks,allcolors=black,breaklinks=true}
\usepackage[margin=2.5cm]{geometry}
\usepackage{graphicx}
\usepackage{float}
\usepackage{times}
\usepackage{url}
\usepackage{enumitem}
\usepackage[dvipsnames]{xcolor}
\usepackage{csquotes}
\usepackage{multicol}
\usepackage{titlesec}
\usepackage[sorting=none]{biblatex}
\usepackage[bitstream-charter]{mathdesign}
\usepackage[T1]{fontenc}
\usepackage{setspace}

\usepackage{tocloft}

\renewcommand\thesection{\arabic{section}}

\usepackage{titlesec}
\titleformat{\section}
  {\huge}{\thesection}{2em}{}
\usepackage{titlesec}
\titleformat{\subsection}
  {\Large}{\thesubsection}{1em}{}

\renewcommand\thesection{\arabic{section}}
\definecolor{custom-yellow}{HTML}{ffea94}
\definecolor{custom-yellow-darker}{HTML}{dbaf00}
\definecolor{custom-green}{HTML}{007849}
\definecolor{custom-blue}{HTML}{0375B4}
\definecolor{custom-blue-darker}{HTML}{035582}
\definecolor{custom-blue-lightest}{HTML}{cdedfe}

\begin{document}

\title{\Huge Toward Trustworthy AI Development:\\Mechanisms for Supporting Verifiable Claims\footnote{Listed authors are those who contributed substantive ideas and/or work to this report. Contributions include writing, research, and/or review for one or more sections; some authors also contributed content via participation in an April 2019 workshop and/or via ongoing discussions. As such, with the exception of the primary/corresponding authors, inclusion as author does not imply endorsement of all aspects of the report.
}}

\author{Miles Brundage$^{1}$\footnote{Miles Brundage (miles@openai.com),  Shahar Avin (sa478@cam.ac.uk),  Jasmine Wang (jasminewang76@gmail.com), Haydn Belfield (hb492@cam.ac.uk), and Gretchen Krueger (gretchen@openai.com) contributed equally and are corresponding authors. Other authors are listed roughly in order of contribution.}, Shahar Avin$^{3,2\dagger}$, Jasmine Wang$^{4,29\dagger}$\footnote{Work conducted in part while at OpenAI.}, Haydn Belfield$^{3,2\dagger}$, Gretchen Krueger$^{1\dagger}$,\\ Gillian Hadfield$^{1,5,30}$, Heidy Khlaaf$^6$, Jingying Yang$^7$, Helen Toner$^8$, Ruth Fong$^9$,\\ Tegan Maharaj$^{4,28}$, Pang Wei Koh$^{10}$, Sara Hooker$^{11}$, Jade Leung$^{12}$, Andrew Trask$^9$,\\ Emma Bluemke$^9$, Jonathan Lebensold$^{4,29}$, Cullen O'Keefe$^1$, Mark Koren$^{13}$, Th\'{e}o Ryffel$^{14}$,\\ JB Rubinovitz$^{15}$, Tamay Besiroglu$^{16}$, Federica Carugati$^{17}$, Jack Clark$^{1}$, Peter Eckersley$^{7}$,\\ Sarah de Haas$^{18}$, Maritza Johnson$^{18}$, Ben Laurie$^{18}$, Alex Ingerman$^{18}$, Igor Krawczuk$^{19}$,\\ Amanda Askell$^{1}$, Rosario Cammarota$^{20}$, Andrew Lohn$^{21}$, David Krueger$^{4,27}$, Charlotte Stix$^{22}$,\\ Peter Henderson$^{10}$, Logan Graham$^{9}$, Carina Prunkl$^{12}$, Bianca Martin$^{1}$, Elizabeth Seger$^{16}$,\\ Noa Zilberman$^{9}$, 
Se\'{a}n \'{O} h\'{E}igeartaigh$^{2,3}$, 
Frens Kroeger$^{23}$, Girish Sastry$^{1}$, Rebecca Kagan$^{8}$,\\ Adrian Weller$^{16,24}$, Brian Tse$^{12,7}$, Elizabeth Barnes$^{1}$, Allan Dafoe$^{12,9}$, Paul Scharre$^{25}$,\\ Ariel Herbert-Voss$^{1}$, Martijn Rasser$^{25}$, Shagun Sodhani$^{4,27}$, Carrick Flynn$^{8}$,\\ Thomas Krendl Gilbert$^{26}$, Lisa Dyer$^{7}$, Saif Khan$^{8}$, Yoshua Bengio$^{4,27}$, Markus Anderljung$^{12}$
\vspace{0.3in}\\
\small{$^1$OpenAI, $^2$Leverhulme Centre for the Future of Intelligence, $^3$Centre for the Study of Existential Risk,}\\\small{$^4$Mila, $^5$University of Toronto, $^6$Adelard, $^7$Partnership on AI, $^8$Center for Security and Emerging Technology,}\\\small{$^9$University of Oxford, $^{10}$Stanford University, $^{11}$Google Brain, $^{12}$Future of Humanity Institute,}\\\small{$^{13}$Stanford Centre for AI Safety, $^{14}$\'{E}cole Normale Sup\'{e}rieure (Paris), $^{15}$Remedy.AI,}\\\small{$^{16}$University of Cambridge, $^{17}$Center for Advanced Study in the Behavioral Sciences,$^{18}$Google Research,}\\\small{$^{19}$\'Ecole Polytechnique F\'ed\'erale de Lausanne, $^{20}$Intel, $^{21}$RAND Corporation,}\\\small{$^{22}$Eindhoven University of Technology, $^{23}$Coventry University, $^{24}$Alan Turing Institute,}\\\small{$^{25}$Center for a New American Security, $^{26}$University of California, Berkeley,}\\\small{$^{27}$University of Montreal, $^{28}$Montreal Polytechnic, $^{29}$McGill University,}\\\small{ $^{30}$Schwartz Reisman Institute for Technology and Society}}

\date{April 2020}

\maketitle
\thispagestyle{empty}

\newpage 

{\begin{spacing}{0.9}
\pagenumbering{gobble}
\tableofcontents
\end{spacing}}

\newpage 
\section*{Executive Summary}\label{Executive Summary}
\pagenumbering{arabic}
\setcounter{page}{1}
\phantomsection\addcontentsline{toc}{section}{Executive Summary}
\vspace{-\baselineskip}
Recent progress in artificial intelligence (AI) has enabled a diverse array of applications across commercial, scientific, and creative domains. With this wave of applications has come a growing awareness of the large-scale impacts of AI systems, and recognition that existing regulations and norms in industry and academia are insufficient to ensure responsible AI development \cite{Askell2019} \cite{Whittaker2018} \cite{Crawford2019}.

Steps have been taken by the AI community to acknowledge and address this insufficiency, including widespread adoption of ethics principles by researchers and technology companies. However, ethics principles are non-binding, and their translation to actions is often not obvious. Furthermore, those outside a given organization are often ill-equipped to assess whether an AI developer's actions are consistent with their stated principles. Nor are they able to hold developers to account when principles and behavior diverge, fueling accusations of "ethics washing" \cite{Gibney2020}.  In order for AI developers to earn trust from system users, customers, civil society, governments, and other stakeholders that they are building AI responsibly, there is a need to move beyond principles to a focus on mechanisms for demonstrating responsible behavior \cite{Mittelstadt2019}. Making and assessing verifiable claims, to which developers can be held accountable, is one crucial step in this direction.

With the ability to make precise claims for which evidence can be brought to bear, AI developers can more readily demonstrate responsible behavior to regulators, the public, and one another. Greater verifiability of claims about AI development would help enable more effective oversight and reduce pressure to cut corners for the sake of gaining a competitive edge \cite{Askell2019}. Conversely, without the capacity to verify claims made by AI developers, those using or affected by AI systems are more likely to be put at risk by potentially ambiguous, misleading, or false claims. 

This report suggests various steps that different stakeholders in AI development can take to make it easier to verify claims about AI development, with a focus on providing evidence about the safety, security, fairness, and privacy protection of AI systems. Implementation of such mechanisms can help make progress on the multifaceted problem of ensuring that AI development is conducted in a trustworthy fashion.\footnote{The capacity to verify claims made by developers, on its own, would be insufficient to ensure responsible AI development. Not all important claims admit verification, and there is also a need for oversight agencies such as governments and standards organizations to align developers' incentives with the public interest.} The mechanisms outlined in this report deal with questions that various parties involved in AI development might face, such as:
\vspace{-\baselineskip}
\begin{itemize}
    \item Can I (as a user) verify the claims made about the level of privacy protection guaranteed by a new AI system I'd like to use for machine translation of sensitive documents? 
    \vspace{-0.3\baselineskip}
    \item Can I (as a regulator) trace the steps that led to an accident caused by an autonomous vehicle? Against what standards should an autonomous vehicle company's safety claims be compared? 
    \vspace{-0.3\baselineskip}
    \item Can I (as an academic) conduct impartial research on the impacts associated with large-scale AI systems when I lack the computing resources of industry? 
    \vspace{-0.3\baselineskip}
    \item Can I (as an AI developer) verify that my competitors in a given area of AI development will follow best practices rather than cut corners to gain an advantage?
\end{itemize}
    \vspace{-0.5\baselineskip}
    
Even AI developers who have the desire and/or incentives to make concrete, verifiable claims may not be equipped with the appropriate mechanisms to do so. The AI development community needs a robust "toolbox" of mechanisms to support the verification of claims about AI systems and development processes.

This problem framing led some of the authors of this report to hold a workshop in April 2019, aimed at expanding the toolbox of mechanisms for making and assessing verifiable claims about AI development.\footnote{See \hyperref[subsec:process]{\textbf{Appendix I, "Workshop and Report Writing Process."}}} This report builds on the ideas proposed at that workshop. The mechanisms outlined do two things:
    \vspace{-\baselineskip}
\begin{itemize}
    \item They increase the options available to AI developers for substantiating claims they make about AI systems' properties.
    \vspace{-0.3\baselineskip}
    \item They increase the specificity and diversity of demands that can be made of AI developers by other stakeholders such as users, policymakers, and members of civil society.
\end{itemize}

Each mechanism and associated recommendation discussed in this report addresses a specific gap preventing effective assessment of developers' claims today. Some of these mechanisms exist and need to be extended or scaled up in some way, and others are novel. The report is intended as an incremental step toward improving the verifiability of claims about AI development.

The report organizes mechanisms under the headings of \textit{Institutions}, \textit{Software}, and \textit{Hardware}, which are three intertwined components of AI systems and development processes. 
\vspace{-\baselineskip}
\begin{itemize}
    \item \hyperref[sec:institutional]{\textbf{Institutional Mechanisms}:} These mechanisms shape or clarify the incentives of people involved in AI development and provide greater visibility into their behavior, including their efforts to ensure that AI systems are safe, secure, fair, and privacy-preserving. Institutional mechanisms play a foundational role in verifiable claims about AI development, since it is \emph{people} who are ultimately responsible for AI development. We focus on \hyperref[subsec:audit]{\textbf{third party auditing}}, to create a robust alternative to self-assessment of claims; \hyperref[subsec:redteam]{\textbf{red teaming exercises}}, to demonstrate AI developers' attention to the ways in which their systems could be misused; \hyperref[subsec:bounties]{\textbf{bias and safety bounties}}, to strengthen incentives to discover and report flaws in AI systems; and \hyperref[subsec:incidents]{\textbf{sharing of AI incidents}}, to improve societal understanding of how AI systems can behave in unexpected or undesired ways.
\vspace{-0.25\baselineskip}

    \item \hyperref[sec:software]{\textbf{Software Mechanisms}:} These mechanisms enable greater understanding and oversight of specific AI systems' properties. We focus on \hyperref[subsec:trails]{\textbf{audit trails}}, to enable accountability for high-stakes AI systems by capturing critical information about the development and deployment process; \hyperref[subsec:interpretabilitysoft]{\textbf{interpretability}}, to foster understanding and scrutiny of AI systems' characteristics; and \textbf{\hyperref[subsec:ppml]{privacy-preserving machine learning}}, to make developers' commitments to privacy protection more robust. 
\vspace{-0.25\baselineskip}
    \item \hyperref[sec:hardware]{\textbf{Hardware Mechanisms}:} Mechanisms related to computing hardware can play a key role in substantiating strong claims about privacy and security, enabling transparency about how an organization's resources are put to use, and influencing who has the resources necessary to verify different claims. We focus on \hyperref[subsec:securehardware]{\textbf{secure hardware for machine learning}}, to increase the verifiability of privacy and security claims; \hyperref[subsec:measurement]{\textbf{high-precision compute measurement}}, to improve the value and comparability of claims about computing power usage; and \hyperref[subsec:support]{\textbf{compute support for academia}}, to improve the ability of those outside of industry to evaluate claims about large-scale AI systems.
\end{itemize}
\vspace{-0.5\baselineskip}

Each mechanism provides additional paths to verifying AI developers' commitments to responsible AI development, and has the potential to contribute to a more trustworthy AI ecosystem. The full list of recommendations associated with each mechanism is found on the following page and again at the end of the report.

\newpage
\phantomsection
\subsection*{List of Recommendations}
\addcontentsline{toc}{subsection}{List of Recommendations}

\hyperref[sec:institutional]{\textbf{Institutional Mechanisms and Recommendations}}
\begin{enumerate}
    \item A coalition of stakeholders should create a task force to research options for conducting and funding \hyperref[subsec:audit]{\textbf{third party auditing}} of AI systems.
    \item Organizations developing AI should run \hyperref[subsec:redteam]{\textbf{red teaming exercises}} to explore risks associated with systems they develop, and should share best practices and tools for doing so.
    \item AI developers should pilot \hyperref[subsec:bounties]{\textbf{bias and safety bounties}} for AI systems to strengthen incentives and processes for broad-based scrutiny of AI systems.
    \item AI developers should share more information about \hyperref[subsec:incidents]{\textbf{AI incidents}}, including through collaborative channels.
\end{enumerate}
\hyperref[sec:software]{\textbf{Software Mechanisms and Recommendations}}
\begin{enumerate}[resume]
    \item Standards setting bodies should work with academia and industry to develop \hyperref[subsec:trails]{\textbf{audit trail}} requirements for safety-critical applications of AI systems.
    \item Organizations developing AI and funding bodies should support research into the \hyperref[subsec:interpretabilitysoft]{\textbf{interpretability}} of AI systems, with a focus on supporting risk assessment and auditing.
    \item AI developers should develop, share, and use suites of tools for \hyperref[subsec:ppml]{\textbf{privacy-preserving machine learning}} that include measures of performance against common standards.
\end{enumerate}
\hyperref[sec:hardware]{\textbf{Hardware Mechanisms and Recommendations}}
\begin{enumerate}[resume]
    \item Industry and academia should work together to develop \hyperref[subsec:securehardware]{\textbf{hardware security features}} for AI accelerators or otherwise establish best practices for the use of secure hardware (including secure enclaves on commodity hardware) in machine learning contexts.
    \item One or more AI labs should estimate the computing power involved in a single project in great detail (\hyperref[subsec:measurement]{\textbf{high-precision compute measurement}}), and report on the potential for wider adoption of such methods.
    \item Government funding bodies should substantially increase \hyperref[subsec:support]{\textbf{funding of computing power resources}} for researchers in academia, in order to improve the ability of those researchers to verify claims made by industry.
\end{enumerate}

\newpage 
\section{Introduction}\label{sec:introduction}
\subsection{Motivation}

With rapid technical progress in artificial intelligence (AI)\footnote{We define AI as digital systems that are capable of performing tasks commonly thought to require intelligence, with these tasks typically learned via data and/or experience.} and the spread of AI-based applications over the past several years, there is growing concern about how to ensure that the development and deployment of AI is beneficial -- and not detrimental -- to humanity. In recent years, AI systems have been developed in ways that are inconsistent with the stated values of those developing them. This has led to a rise in concern, research, and activism relating to the impacts of AI systems \cite{Whittaker2018} \cite{Crawford2019}. AI development has raised concerns about amplification of bias \cite{Hao2019}, loss of privacy \cite{UniversityofCalifornia-Berkeley2019}, digital addictions \cite{Alter2017},  social harms associated with facial recognition and criminal risk assessment \cite{Angwin2016}, disinformation \cite{Solaiman2019}, and harmful changes to the quality \cite{Gray2019} and availability of gainful employment \cite{BenediktFrey2013}.

In response to these concerns, a range of stakeholders, including those developing AI systems, have articulated ethics principles to guide responsible AI development. The amount of work undertaken to articulate and debate such principles is encouraging, as is the convergence of many such principles on a set of widely-shared concerns such as safety, security, fairness, and privacy.\footnote{Note, however, that many such principles have been articulated by Western academics and technology company employees, and as such are not necessarily representative of humanity's interests or values as a whole. Further, they are amenable to various interpretations \cite{Whittlestone2019}\cite{Jobin2019} and agreement on them can mask deeper disagreements \cite{Mittelstadt2019}. See also Beijing AI Principles \cite{BeijingAIPrinciples2019} and Zeng et. al. \cite{Zeng2018} for examples of non-Western AI principles.}

However, \textbf{principles are only a first step in the effort to ensure beneficial societal outcomes from AI} \cite{Whittlestone2019}. Indeed, studies \cite{Edelman2019}, surveys \cite{Zhang2019a}, and trends in worker and community organizing \cite{Whittaker2018} \cite{Crawford2019} make clear that large swaths of the public are concerned about the risks of AI development, and do not trust the organizations currently dominating such development to self-regulate effectively. Those potentially affected by AI systems need mechanisms for ensuring responsible development that are more robust than high-level principles. People who get on airplanes don't trust an airline manufacturer because of its PR campaigns about the importance of safety - they trust it because of the accompanying infrastructure of technologies, norms, laws, and institutions for ensuring airline safety.\footnote{Recent commercial airline crashes also serve as a reminder that even seemingly robust versions of such infrastructure are imperfect and in need of constant vigilance.} Similarly, along with the growing explicit adoption of ethics principles to guide AI development, there is mounting skepticism about whether these claims and commitments can be monitored and enforced \cite{Travis2019}.

Policymakers are beginning to enact regulations that more directly constrain AI developers' behavior \cite{Gesley2019}. We believe that analyzing AI development through the lens of verifiable claims can help to inform such efforts. AI developers, regulators, and other actors all need to understand which properties of AI systems and development processes can be credibly demonstrated, through what means, and with what tradeoffs.

We define \textbf{verifiable claims}\footnote{While this report does discuss the technical area of formal verification at several points, and several of our recommendations are based on best practices from the field of information security, the sense in which we use "verifiable" is distinct from how the term is used in those contexts. Unless otherwise specified by the use of the adjective "formal" or other context, this report uses the word verification in a looser sense. Formal verification seeks mathematical proof that a certain technical claim is true with certainty (subject to certain assumptions). In contrast, this report largely focuses on claims that are unlikely to be demonstrated with absolute certainty, but which can be shown to be likely or unlikely to be true through relevant arguments and evidence.}  as falsifiable statements for which evidence and arguments can be brought to bear on the likelihood of those claims being true. While the degree of attainable certainty will vary across different claims and contexts, we hope to show that greater degrees of evidence can be provided for claims about AI development than is typical today. The nature and importance of verifiable claims is discussed in greater depth in \hyperref[subsec:importance]{\textbf{Appendix III}}, and we turn next to considering the types of mechanisms that can make claims verifiable.

\subsection{Institutional, Software, and Hardware Mechanisms}

AI developers today have many possible approaches for increasing the verifiability of their claims. Despite the availability of many mechanisms that could help AI developers demonstrate their claims and help other stakeholders scrutinize their claims, this toolbox has not been well articulated to date. 

We view AI development processes as sociotechnical systems,\footnote{Broadly, a sociotechnical system is one whose "core interface consists of the relations between a nonhuman system and a human system", rather than the components of those systems in isolation. See Trist \cite{Trist1981}.} with institutions, software, and hardware all potentially supporting (or detracting from) the verifiability of claims about AI development. AI developers can make claims about, or take actions related to, each of these three interrelated pillars of AI development. 

In some cases, adopting one of these mechanisms can increase the verifiability of one's own claims, whereas in other cases the impact on trust is more indirect (i.e., a mechanism implemented by one actor enabling greater scrutiny of other actors). As such, collaboration across sectors and organizations will be critical in order to build an ecosystem in which claims about responsible AI development can be verified. 

\begin{itemize}
    \item Institutional mechanisms largely pertain to \textit{values, incentives, and accountability}. Institutional mechanisms shape or clarify the incentives of people involved in AI development and provide greater visibility into their behavior, including their efforts to ensure that AI systems are safe, secure, fair, and privacy-preserving. These mechanisms can also create or strengthen channels for holding AI developers accountable for harms associated with AI development. In this report, we provide an overview of some such mechanisms, and then discuss \textbf{third party auditing}, \textbf{red team exercises}, \textbf{safety and bias bounties}, and \textbf{sharing of AI incidents} in more detail.
    \item Software mechanisms largely pertain to \textit{specific AI systems and their properties}. Software mechanisms can be used to provide evidence for both formal and informal claims regarding the properties of specific AI systems, enabling greater understanding and oversight. The software mechanisms we highlight below are \textbf{audit trails}, \textbf{interpretability}, and \textbf{privacy-preserving machine learning}.
    \item Hardware mechanisms largely pertain to \textit{physical computational resources and their properties}. Hardware mechanisms can support verifiable claims by providing greater assurance regarding the privacy and security of AI systems, and can be used to substantiate claims about how an organization is using their general-purpose computing capabilities. Further, the distribution of resources across different actors can influence the types of AI systems that are developed and which actors are capable of assessing other actors' claims (including by reproducing them). The hardware mechanisms we focus on in this report are \textbf{hardware security features for machine learning}, \textbf{high-precision compute measurement}, and \textbf{computing power support for academia}.
\end{itemize}

\subsection{Scope and Limitations}

This report focuses on a particular aspect of trustworthy AI development: the extent to which organizations developing AI systems can and do make verifiable claims about the AI systems they build, and the ability of other parties to assess those claims. Given the backgrounds of the authors, the report focuses in particular on mechanisms for demonstrating claims about AI systems being safe, secure, fair, and/or privacy-preserving, without implying that those are the only sorts of claims that need to be verified. 

We devote particular attention to mechanisms\footnote{We use the term mechanism generically to refer to processes, systems, or approaches for providing or generating evidence about behavior.} that the authors have expertise in and for which concrete and beneficial next steps were identified at an April 2019 workshop. These are not the only mechanisms relevant to verifiable claims; we survey some others at the beginning of each section, and expect that further useful mechanisms have yet to be identified. 

Making verifiable claims is part of, but not equivalent to, trustworthy AI development, broadly defined. An AI developer might also be more or less trustworthy based on the particular values they espouse, the extent to which they engage affected communities in their decision-making, or the extent of recourse that they provide to external parties who are affected by their actions. Additionally, the actions of AI developers, which we focus on, are not all that matters for trustworthy AI development--the existence and enforcement of relevant laws matters greatly, for example. 

\hyperref[subsec:process]{\textbf{Appendix I}} discusses the reasons for the report's scope in more detail, and \hyperref[subsec:terms]{\textbf{Appendix II}} discusses the relationship between different definitions of trust and verifiable claims. When we use the term "trust" as a verb in the report, we mean that one party (party A) gains confidence in the reliability of another party's claims (party B) based on evidence provided about the accuracy of those claims or related ones. We also make reference to this claim-oriented sense of trust when we discuss actors "earning" trust, (providing evidence for claims made), or being "trustworthy" (routinely providing sufficient evidence for claims made). This use of language is intended to concisely reference an important dimension of trustworthy AI development, and is not meant to imply that verifiable claims are sufficient for attaining trustworthy AI development. 

\subsection{Outline of the Report}

The next three sections of the report, \hyperref[sec:institutional]{\textbf{Institutional Mechanisms and Recommendations}}, \hyperref[sec:software]{\textbf{Software Mechanisms and Recommendations}}, and \hyperref[sec:hardware]{\textbf{Hardware Mechanisms and Recommendations}}, each begin with a survey of mechanisms relevant to that category. Each section then highlights several mechanisms that we consider especially promising. We are uncertain which claims are most important to verify in the context of AI development, but strongly suspect that some combination of the mechanisms we outline in this report are needed to craft an AI ecosystem in which responsible AI development can flourish.

The way we articulate the case for each mechanism is \textit{problem-centric}: each mechanism helps address a potential barrier to claim verification identified by the authors. Depending on the case, the recommendations associated with each mechanism are aimed at implementing a mechanism for the first time, researching it, scaling it up, or extending it in some way.

The \hyperref[sec:conclusion]{\textbf{Conclusion}} puts the report in context, discusses some important caveats, and reflects on next steps.

The \hyperref[subsec:process]{\textbf{Appendices}} provide important context, supporting material, and supplemental analysis. \hyperref[subsec:process]{\textbf{Appendix I}} provides background on the workshop and the process that went into writing the report; \hyperref[subsec:terms]{\textbf{Appendix II}} serves as a glossary and discussion of key terms used in the report; \hyperref[subsec:importance]{\textbf{Appendix III}} discusses the nature and importance of verifiable claims; \hyperref[subsec:arms]{\textbf{Appendix IV}} discusses the importance of verifiable claims in the context of arms control; \hyperref[subsec:laws]{\textbf{Appendix V}} provides context on antitrust law as it relates to cooperation among AI developers on responsible AI development; and \hyperref[subsec:analysis]{\textbf{Appendix VI}} offers supplemental analysis of several mechanisms.

\newpage 
\section{Institutional Mechanisms and Recommendations}\label{sec:institutional}

"Institutional mechanisms" are processes that shape or clarify the incentives of the people involved in AI development, make their behavior more transparent, or enable accountability for their behavior. Institutional mechanisms help to ensure that individuals or organizations making claims regarding AI development are incentivized to be diligent in developing AI responsibly and that other stakeholders can verify that behavior. Institutions\footnote{\textbf{Institutions} may be formal and public institutions, such as: laws, courts, and regulatory agencies; private formal arrangements between parties, such as contracts; interorganizational structures such as industry associations, strategic alliances, partnerships, coalitions, joint ventures, and research consortia. Institutions may also be informal norms and practices that prescribe behaviors in particular contexts; or third party organizations, such as professional bodies and academic institutions.} can shape incentives or constrain behavior in various ways. 

Several clusters of existing institutional mechanisms are relevant to responsible AI development, and we characterize some of their roles and limitations below. These provide a foundation for the subsequent, more detailed discussion of several mechanisms and associated recommendations. Specifically, we provide an overview of some existing institutional mechanisms that have the following functions:
\begin{itemize}
    \item Clarifying organizational goals and values;
    \item Increasing transparency regarding AI development processes;
    \item Creating incentives for developers to act in ways that are responsible; and
    \item Fostering exchange of information among developers.
\end{itemize}

Institutional mechanisms can \textbf{help clarify an organization's goals and values}, which in turn can provide a basis for evaluating their claims. These statements of goals and values--which can also be viewed as (high level) claims in the framework discussed here--can help to contextualize the actions an organization takes and lay the foundation for others (shareholders, employees, civil society organizations, governments, etc.) to monitor and evaluate behavior. Over 80 AI organizations \cite{Mittelstadt2019}, including technology companies such as Google \cite{Pichai2018}, OpenAI \cite{OpenAI2018}, and Microsoft \cite{Microsofta} have publicly stated the principles they will follow in developing AI. Codes of ethics or conduct are far from sufficient, since they are typically abstracted away from particular cases and are not reliably enforced, but they can be valuable by establishing criteria that a developer concedes are appropriate for evaluating its behavior. 

The creation and public announcement of a \textit{code of ethics} proclaims an organization's commitment to ethical conduct both externally to the wider public, as well as internally to its employees, boards, and shareholders. \textit{Codes of conduct} differ from codes of ethics in that they contain a set of concrete behavioral standards.\footnote{Many organizations use the terms synonymously. The specificity of codes of ethics can vary, and more specific (i.e., action-guiding) codes of ethics (i.e. those equivalent to codes of conduct) can be better for earning trust because they are more falsifiable. Additionally, the form and content of these mechanisms can evolve over time--consider, e.g., Google's AI Principles, which have been incrementally supplemented with more concrete guidance in particular areas.}

Institutional mechanisms \textbf{can increase transparency regarding an organization's AI development processes} in order to permit others to more easily verify compliance with appropriate norms, regulations, or agreements. Improved transparency may reveal the extent to which actions taken by an AI developer are consistent with their declared intentions and goals. The more reliable, timely, and complete the institutional measures to enhance transparency are, the more assurance may be provided. 

Transparency measures could be undertaken on a voluntary basis or as part of an agreed framework involving relevant parties (such as a consortium of AI developers, interested non-profits, or policymakers).  For example, \textit{algorithmic impact assessments} are intended to support affected communities and stakeholders in assessing AI and other automated decision systems \cite{Whittaker2018}. The Canadian government, for example, has centered AIAs in its Directive on Automated Decision-Making \cite{GovernmentofCanada} \cite{TreasuryBoardofCanadaSecretariat2019}. Another path toward greater transparency around AI development involves increasing the extent and quality of \emph{documentation} for AI systems. Such documentation can help foster informed and safe use of AI systems by providing information about AI systems' biases and other attributes \cite{Raji2019}\cite{Mitchell2018}\cite{Gebru2018}.

Institutional mechanisms can \textbf{create incentives for organizations to act in ways that are responsible}. Incentives can be created within an organization or externally, and they can operate at an organizational or an individual level. The incentives facing an actor can provide evidence regarding how that actor will behave in the future, potentially bolstering the credibility of related claims. To modify incentives at an organizational level, organizations can choose to \textit{adopt different organizational structures} (such as benefit corporations) or take on \textit{legally binding intra-organizational commitments}. For example, organizations could credibly commit to distributing the benefits of AI broadly through a legal commitment that shifts fiduciary duties.\footnote{The Windfall Clause \cite{OKeefe2019} is one proposal along these lines, and involves an ex ante commitment by AI firms to donate a significant amount of any eventual extremely large profits.} 

Institutional commitments to such steps could make a particular organization's financial incentives more clearly aligned with the public interest. To the extent that commitments to responsible AI development and distribution of benefits are widely implemented, AI developers would stand to benefit from each others' success, potentially\footnote{The global nature of AI development, and the national nature of much relevant regulation, is a key complicating factor.} reducing incentives to race against one another \cite{Askell2019}. And critically, government regulations such as the General Data Protection Regulation (GDPR) enacted by the European Union shift developer incentives by imposing penalties on developers that do not adequately protect privacy or provide recourse for algorithmic decision-making.

Finally, institutional mechanisms can \textbf{foster exchange of information between developers}. To avoid "races to the bottom" in AI development, AI developers can exchange lessons learned and demonstrate their compliance with relevant norms to one another. Multilateral fora (in addition to bilateral conversations between organizations) provide opportunities for discussion and repeated interaction, increasing transparency and interpersonal understanding. Voluntary membership organizations with stricter rules and norms have been implemented in other industries and might also be a useful model for AI developers \cite{Buchanan1965}.\footnote{See for example the norms set and enforced by the European Telecommunications Standards Institute (ETSI). These norms have real "teeth," such as the obligation for designated holders of Standard Essential Patents to license on Fair, Reasonable and Non-discriminatory (FRAND) terms. Breach of FRAND could give rise to a breach of contract claim as well as constitute a breach of antitrust law \cite{ETSI}. Voluntary standards for consumer products, such as those associated with Fairtrade and Organic labels, are also potentially relevant precedents \cite{Trade2010}.} 

Steps in the direction of robust information exchange between AI developers include the creation of consensus around important priorities such as safety, security, privacy, and fairness;\footnote{An example of such an effort is the Asilomar AI Principles \cite{FutureofLifeInstitute2017}.} participation in multi-stakeholder fora such as the Partnership on Artificial Intelligence to Benefit People and Society (PAI), the Association for Computing Machinery (ACM), the Institute of Electrical and Electronics Engineers (IEEE), the International Telecommunications Union (ITU), and the International Standards Organization (ISO); and clear identification of roles or offices within organizations who are responsible for maintaining and deepening interorganizational communication \cite{Solaiman2019}.\footnote{Though note competitors sharing commercially sensitive, non-public information (such as strategic plans or R\&D plans) could raise antitrust concerns. It is therefore important to have the right antitrust governance structures and procedures in place (i.e., setting out exactly what can and cannot be shared). See \hyperref[subsec:laws]{\textbf{Appendix V}}. } 

It is also important to examine the incentives (and disincentives) for free flow of information within an organization. Employees within organizations developing AI systems can play an important role in identifying unethical or unsafe practices. For this to succeed, employees must be well-informed about the scope of AI development efforts within their organization and be comfortable raising their concerns, and such concerns need to be taken seriously by management.\footnote{Recent revelations regarding the culture of engineering and management at Boeing highlight the urgency of this issue \cite{Pasztor2020}.} Policies (whether governmental or organizational) that help ensure safe channels for expressing concerns are thus key foundations for verifying claims about AI development being conducted responsibly.

The subsections below each introduce and explore a mechanism with the potential for improving the verifiability of claims in AI development: \hyperref[subsec:audit]{\textbf{third party auditing}}, \hyperref[subsec:redteam]{\textbf{red team exercises}}, \hyperref[subsec:bounties]{\textbf{bias and safety bounties}}, and \hyperref[subsec:incidents]{\textbf{sharing of AI incidents}}. In each case, the subsections below begin by discussing a problem which motivates exploration of that mechanism, followed by a recommendation for improving or applying that mechanism.

\newpage 

\subsection{Third Party Auditing}\label{subsec:audit}
\setlength{\fboxsep}{6pt}
\textbf{Problem:}

\vspace{-0.5\baselineskip}
\fbox{%
    \parbox{0.97\textwidth}{%
        \textbf{The process of AI development is often opaque to those outside a given organization, and various barriers make it challenging for third parties to verify the claims being made by a developer. As a result, claims about system attributes may not be easily verified.}
    }%
}

AI developers have justifiable concerns about being transparent with information concerning commercial secrets, personal information, or AI systems that could be misused; however, problems arise when these concerns incentivize them to evade scrutiny. Third party auditors can be given privileged and secured access to this private information, and they can be tasked with assessing whether safety, security, privacy, and fairness-related claims made by the AI developer are accurate. 

Auditing is a structured process by which an organization's present or past behavior is assessed for consistency with relevant principles, regulations, or norms. Auditing has promoted consistency and accountability in industries outside of AI such as finance and air travel. In each case, auditing is tailored to the evolving nature of the industry in question.\footnote{See  Raji and Smart et al. \cite{Raji2020} for a discussion of some lessons for AI from auditing in other industries.} Recently, auditing has gained traction as a potential paradigm for assessing whether AI development was conducted in a manner consistent with the stated principles of an organization, with valuable work focused on designing internal auditing processes (i.e. those in which the auditors are also employed by the organization being audited) \cite{Raji2020}.

Third party auditing is a form of auditing conducted by an external and independent auditor, rather than the organization being audited, and can help address concerns about the incentives for accuracy in self-reporting. Provided that they have sufficient information about the activities of an AI system, independent auditors with strong reputational and professional incentives for truthfulness can help verify claims about AI development.

Auditing could take at least four quite different forms, and likely further variations are possible: auditing by an independent body with government-backed policing and sanctioning power; auditing that occurs entirely within the context of a government, though with multiple agencies involved \cite{Engstrom2019}; auditing by a private expert organization or some ensemble of such organizations; and internal auditing followed by public disclosure of (some subset of) the results.\footnote{Model cards for model reporting \cite{Mitchell2018} and data sheets for datasets \cite{Gebru2018} reveal information about AI systems publicly, and future work in third party auditing could build on such tools, as advocated by Raji and Smart et al. \cite{Raji2020}.} As commonly occurs in other contexts, the results produced by  independent auditors might be made publicly available, to increase confidence in the propriety of the auditing process.\footnote{Consumer Reports, originally founded as the Consumers Union in 1936, is one model for an independent, third party organization that performs similar functions for products that can affect the health, well-being, and safety of the people using those products. (https://www.consumerreports.org/cro/about-us/what-we-do/research-and-testing/index.htm).}

Techniques and best practices have not yet been established for auditing AI systems. Outside of AI, however, there are well-developed frameworks on which to build. Outcomes- or claim-based "assurance frameworks" such as the Claims-Arguments-Evidence framework (CAE) and Goal Structuring Notation (GSN) are already in wide use in safety-critical auditing contexts.\footnote{See \hyperref[subsec:importance]{\textbf{Appendix III}} for further discussion of claim-based frameworks for auditing.}  By allowing different types of arguments and evidence to be used appropriately by auditors, these frameworks provide considerable flexibility in how high-level claims are substantiated, a needed feature given the wide ranging and fast-evolving societal challenges posed by AI.

Possible aspects of AI systems that could be independently audited include the level of privacy protection guaranteed, the extent to (and methods by) which the AI systems were tested for safety, security or ethical concerns, and the sources of data, labor, and other resources used. Third party auditing could be applicable to a wide range of AI applications, as well. Safety-critical AI systems such as autonomous vehicles and medical AI systems, for example, could be audited for safety and security. Such audits could confirm or refute the accuracy of previous claims made by developers, or compare their efforts against an independent set of standards for safety and security. As another example, search engines and recommendation systems could be independently audited for harmful biases. 

Third party auditors should be held accountable by government, civil society, and other stakeholders to ensure that strong incentives exist to act accurately and fairly. Reputational considerations help to ensure auditing integrity in the case of financial accounting, where firms prefer to engage with credible auditors \cite{Barton2005}. Alternatively, a licensing system could be implemented in which auditors undergo a standard training process in order to become a licensed AI system auditor. However, given the variety of methods and applications in the field of AI, it is not obvious whether auditor licensing is a feasible option for the industry: perhaps a narrower form of licensing would be helpful (e.g., a subset of AI such as adversarial machine learning).

Auditing imposes costs (financial and otherwise) that must be weighed against its value. Even if auditing is broadly societally beneficial and non-financial costs (e.g., to intellectual property) are managed, the financial costs will need to be borne by someone (auditees, large actors in the industry, taxpayers, etc.), raising the question of how to initiate a self-sustaining process by which third party auditing could mature and scale. However, if done well, third party auditing could strengthen the ability of stakeholders in the AI ecosystem to make and assess verifiable claims. And notably, the insights gained from third party auditing could be shared widely, potentially benefiting stakeholders even in countries with different regulatory approaches for AI.

\textbf{Recommendation: A coalition of stakeholders should create a task force to research options for conducting and funding third party auditing of AI systems.}

AI developers and other stakeholders (such as civil society organizations and policymakers) should collaboratively explore the challenges associated with third party auditing. A task force focused on this issue could explore appropriate initial domains/applications to audit, devise approaches for handling sensitive intellectual property, and balance the need for standardization with the need for flexibility as AI technology evolves.\footnote{This list is not exhaustive - see, e.g., \cite{Carrier2019}, \cite{Hempel2018}, and \cite{Etzioni2019} for related discussions.} Collaborative research into this domain seems especially promising given that the same auditing process could be used across labs and countries. As research in these areas evolves, so too will auditing processes--one might thus think of auditing as a "meta-mechanism" which could involve assessing the quality of other efforts discussed in this report such as red teaming. 

One way that third party auditing could connect to government policies, and be funded, is via a "regulatory market" \cite{Clark2019}. In a regulatory market for AI, a government would establish high-level outcomes to be achieved from regulation of AI (e.g., achievement of a certain level of safety in an industry) and then create or support private sector entities or other organizations that compete in order to design and implement the precise technical oversight required to achieve those outcomes.\footnote{Examples of such entities include EXIDA, the UK Office of Nuclear Regulation, and the private company Adelard.} Regardless of whether such an approach is pursued, third party auditing by private actors should be viewed as a complement to, rather than a substitute, for governmental regulation. And regardless of the entity conducting oversight of AI developers, in any case there will be a need to grapple with difficult challenges such as the treatment of proprietary data. 

\newpage 

\subsection{Red Team Exercises}\label{subsec:redteam}
\setlength{\fboxsep}{6pt}
\textbf{Problem:}

\vspace{-0.5\baselineskip}
\fbox{%
    \parbox{0.97\textwidth}{%
        \textbf{It is difficult for AI developers to address the "unknown unknowns" associated with AI systems, including limitations and risks that might be exploited by malicious actors. Further, existing red teaming approaches are insufficient for addressing these concerns in the AI context.}
    }%
}

In order for AI developers to make verifiable claims about their AI systems being safe or secure, they need processes for surfacing and addressing potential safety and security risks. Practices such as red teaming exercises help organizations to discover their own limitations and vulnerabilities as well as those of the AI systems they develop, and to approach them holistically, in a way that takes into account the larger environment in which they are operating.\footnote{Red teaming could be aimed at assessing various properties of AI systems, though we focus on safety and security in this subsection given the expertise of the authors who contributed to it.}  

A red team exercise is a structured effort to find flaws and vulnerabilities in a plan, organization, or technical system, often performed by dedicated "red teams" that seek to adopt an attacker's mindset and methods. In domains such as computer security, red teams are routinely tasked with emulating attackers in order to find flaws and vulnerabilities in organizations and their systems. Discoveries made by red teams allow organizations to improve security and system integrity before and during deployment. Knowledge that a lab has a red team can potentially improve the trustworthiness of an organization with respect to their safety and security claims, at least to the extent that effective red teaming practices exist and are demonstrably employed.

As indicated by the number of cases in which AI systems cause or threaten to cause harm, developers of an AI system often fail to anticipate the potential risks associated with technical systems they develop. These risks include both inadvertent failures and deliberate misuse. Those not involved in the development of a particular system may be able to more easily adopt and practice an attacker's skillset. A growing number of industry labs have dedicated red teams, although best practices for such efforts are generally in their early stages.\footnote{For an example of early efforts related to this, see Marshall et al., "Threat Modeling AI/ML Systems and Dependencies" \cite{Marshall2019}} There is a need for experimentation both within and across organizations in order to move red teaming in AI forward, especially since few AI developers have expertise in relevant areas such as threat modeling and adversarial machine learning \cite{Herbert-Voss2019}.

AI systems and infrastructure vary substantially in terms of their properties and risks, making in-house red-teaming expertise valuable for organizations with sufficient resources. However, it would also be beneficial to experiment with the formation of \textit{a community of AI red teaming professionals} that draws together individuals from different organizations and backgrounds, specifically focused on some subset of AI (versus AI in general) that is relatively well-defined and relevant across multiple organizations.\footnote{In the context of language models, for example, 2019 saw a degree of communication and coordination across AI developers to assess the relative risks of different language understanding and generation systems \cite{Solaiman2019}. Adversarial machine learning, too, is an area with substantial sharing of lessons across organizations, though it is not obvious whether a shared red team focused on this would be too broad.} A community of red teaming professionals could take actions such as publish best practices, collectively analyze particular case studies, organize workshops on emerging issues, or advocate for policies that would enable red teaming to be more effective.

Doing red teaming in a more collaborative fashion, as a community of focused professionals across organizations, has several potential benefits:
\begin{itemize}
    \item Participants in such a community would gain useful, broad knowledge about the AI ecosystem, allowing them to identify common attack vectors and make periodic ecosystem-wide recommendations to organizations that are not directly participating in the core community;
    \item Collaborative red teaming distributes the costs for such a team across AI developers, allowing those who otherwise may not have utilized a red team of similarly high quality or one at all to access its benefits (e.g., smaller organizations with less resources);
    \item Greater collaboration could facilitate sharing of information about security-related AI incidents.\footnote{This has a precedent from cybersecurity; MITRE's ATT\&CK is a globally accessible knowledge base of adversary tactics and techniques based on real-world observations, which serves as a foundation for development of more specific threat models and methodologies to improve cybersecurity (https://attack.mitre.org/).}
\end{itemize}

\textbf{Recommendation: Organizations developing AI should run red teaming exercises to explore risks associated with systems they develop, and should share best practices and tools for doing so.}

Two critical questions that would need to be answered in the context of forming a more cohesive AI red teaming community are: what is the appropriate scope of such a group, and how will proprietary information be handled?\footnote{These practical questions are not exhaustive, and even addressing them effectively might not suffice to ensure that collaborative red teaming is beneficial. For example, one potential failure mode is if collaborative red teaming fostered excessive homogeneity in the red teaming approaches used, contributing to a false sense of security in cases where that approach is insufficient.} The two questions are related. Particularly competitive contexts (e.g., autonomous vehicles) might be simultaneously very appealing and challenging: multiple parties stand to gain from pooling of insights, but collaborative red teaming in such contexts is also challenging because of intellectual property and security concerns. 

As an alternative to or supplement to explicitly collaborative red teaming, organizations building AI technologies should establish shared resources and outlets for sharing relevant non-proprietary information. The subsection on \hyperref[subsec:incidents]{\textbf{sharing of AI incidents}} also discusses some potential innovations that could alleviate concerns around sharing proprietary information.

\newpage 

\subsection{Bias and Safety Bounties}\label{subsec:bounties}
\setlength{\fboxsep}{6pt}
\textbf{Problem:}

\vspace{-0.5\baselineskip}
\fbox{%
    \parbox{0.97\textwidth}{%
        \textbf{There is too little incentive, and no formal process, for individuals unaffiliated with a particular AI developer to seek out and report problems of AI bias and safety. As a result, broad-based scrutiny of AI systems for these properties is relatively rare.}
    }%
}

"Bug bounty" programs have been popularized in the information security industry as a way to compensate individuals for recognizing and reporting bugs, especially those related to exploits and vulnerabilities \cite{Hackerone2017}. Bug bounties provide a legal and compelling way to report bugs directly to the institutions affected, rather than exposing the bugs publicly or selling the bugs to others. Typically, bug bounties involve an articulation of the scale and severity of the bugs in order to determine appropriate compensation. 

While efforts such as red teaming are focused on bringing internal resources to bear on identifying risks associated with AI systems, bounty programs give outside individuals a method for raising concerns about specific AI systems in a formalized way. Bounties provide one way to increase the amount of scrutiny applied to AI systems, increasing the likelihood of claims about those systems being verified or refuted. 

Bias\footnote{For an earlier exploration of bias bounties by one of the report authors, see Rubinovitz \cite{Rubinovitz2018}.} and safety bounties would extend the bug bounty concept to AI, and could complement existing efforts to better document datasets and models for their performance limitations and other properties.\footnote{For example, model cards for model reporting \cite{Mitchell2018} and datasheets for datasets \cite{Gebru2018} are recently developed means of documenting AI releases, and such documentation could be extended with publicly listed incentives for finding new forms of problematic behavior not captured in that documentation.} We focus here on bounties for discovering bias and safety issues in AI systems as a starting point for analysis and experimentation, but note that bounties for other properties (such as security, privacy protection, or interpretability) could also be explored.\footnote{Bounties for finding issues with datasets used for training AI systems could also be considered, though we focus on trained AI systems and code as starting points.}

While some instances of bias are easier to identify, others can only be uncovered with significant analysis and resources. For example, Ziad Obermeyer et al. uncovered racial bias in a widely used algorithm affecting millions of patients \cite{Obermeyer447}. There have also been several instances of consumers with no direct access to AI institutions using social media and the press to draw attention to problems with AI \cite{Telford2019}. To date, investigative journalists and civil society organizations have played key roles in surfacing different biases in deployed AI systems. If companies were more open earlier in the development process about possible faults, and if users were able to raise (and be compensated for raising) concerns about AI to institutions, users might report them directly instead of seeking recourse in the court of public opinion.\footnote{We note that many millions of dollars have been paid to date via bug bounty programs in the computer security domain, providing some evidence for this hypothesis. However, bug bounties are not a panacea and recourse to the public is also appropriate in some cases.} 

In addition to bias, bounties could also add value in the context of claims about AI safety. Algorithms or models that are purported to have favorable safety properties, such as enabling safe exploration or robustness to distributional shifts \cite{Amodei2016}, could be scrutinized via bounty programs. To date, more attention has been paid to documentation of models for bias properties than safety properties,\footnote{We also note that the challenge of avoiding harmful biases is sometimes framed as a subset of safety, though for the purposes of this discussion, little hinges on this terminological issue. We distinguish the two in the title of this section in order to call attention to the unique properties of different types of bounties.} though in both cases, benchmarks remain in an early state. Improved safety metrics could increase the comparability of bounty programs and the overall robustness of the bounty ecosystem; however, there should also be means of reporting issues that are not well captured by existing metrics. 

Note that bounties are not sufficient for ensuring that a system is safe, secure, or fair, and it is important to avoid creating perverse incentives (e.g., encouraging work on poorly-specified bounties and thereby negatively affecting talent pipelines) \cite{Ashford2018}. Some system properties can be difficult to discover even with bounties, and the bounty hunting community might be too small to create strong assurances. However, relative to the status quo, bounties might increase the amount of scrutiny applied to AI systems. 

\textbf{Recommendation: AI developers should pilot bias and safety bounties for AI systems to strengthen incentives and processes for broad-based scrutiny of AI systems.}

Issues to be addressed in setting up such a bounty program include \cite{Rubinovitz2018}:
\begin{itemize}
    \item Setting compensation rates for different scales/severities of issues discovered;
    \item Determining processes for soliciting and evaluating bounty submissions;
    \item Developing processes for disclosing issues discovered via such bounties in a timely fashion;\footnote{Note that we specifically consider public bounty programs here, though instances of private bounty programs also exist in the computer security community. Even in the event of a publicly advertised bounty, however, submissions may be private, and as such there is a need for explicit policies for handling submissions in a timely and legitimate fashion--otherwise such programs will provide little assurance.}
    \item Designing appropriate interfaces for reporting of bias and safety problems in the context of deployed AI systems;
    \item Defining processes for handling reported bugs and deploying fixes;
    \item Avoiding creation of perverse incentives.
\end{itemize}

There is not a perfect analogy between discovering and addressing traditional computer security vulnerabilities, on the one hand, and identifying and addressing limitations in AI systems, on the other. Work is thus needed to explore the factors listed above in order to adapt the bug bounty concept to the context of AI development. The computer security community has developed norms (though not a consensus) regarding how to address "zero day" vulnerabilities,\footnote{A zero-day vulnerability is a security vulnerability that is unknown to the developers of the system and other affected parties, giving them "zero days" to mitigate the issue if the vulnerability were to immediately become widely known. The computer security community features a range of views on appropriate responses to zero-days, with a common approach being to provide a finite period for the vendor to respond to notification of the vulnerability before the discoverer goes public.} but no comparable norms yet exist in the AI community. 

There may be a need for distinct approaches to different types of vulnerabilities and associated bounties, depending on factors such as the potential for remediation of the issue and the stakes associated with the AI system. Bias might be treated differently from safety issues such as unsafe exploration, as these have distinct causes, risks, and remediation steps. In some contexts, a bounty might be paid for information even if there is no ready fix to the identified issue, because providing accurate documentation to system users is valuable in and of itself and there is often no pretense of AI systems being fully robust. In other cases, more care will be needed in responding to the identified issue, such as when a model is widely used in deployed products and services.

\newpage 

\subsection{Sharing of AI Incidents}\label{subsec:incidents}
\setlength{\fboxsep}{6pt}
\textbf{Problem:}

\vspace{-0.5\baselineskip}
\fbox{%
    \parbox{0.97\textwidth}{%
        \textbf{Claims about AI systems can be scrutinized more effectively if there is common knowledge of the potential risks of such systems. However, cases of desired or unexpected behavior by AI systems are infrequently shared since it is costly to do unilaterally.}
    }%
}

Organizations can share AI "incidents," or cases of undesired or unexpected behavior by an AI system that causes or could cause harm, by publishing case studies about these incidents from which others can learn. This can be accompanied by information about how they have worked to prevent future incidents based on their own and others' experiences. 

By default, organizations developing AI have an incentive to primarily or exclusively report positive outcomes associated with their work rather than incidents. As a result, a skewed image is given to the public, regulators, and users about the potential risks associated with AI development. 

The sharing of AI incidents can improve the verifiability of claims in AI development by highlighting risks that might not have otherwise been considered by certain actors. Knowledge of these risks, in turn, can then be used to inform questions posed to AI developers, increasing the effectiveness of external scrutiny. Incident sharing can also (over time, if used regularly) provide evidence that incidents are found and acknowledged by particular organizations, though additional mechanisms would be needed to demonstrate the completeness of such sharing. 

AI incidents can include those that are publicly known and transparent, publicly known and anonymized, privately known and anonymized, or privately known and transparent. The Partnership on AI has begun building an AI incident-sharing database, called the AI Incident Database.\footnote{See Partnership on AI's AI Incident Registry as an example (http://aiid.partnershiponai.org/). A related resource is a list called Awful AI, which is intended to raise awareness of misuses of AI and to spur discussion around contestational research and tech projects \cite{Dao}. A separate list summarizes various cases in which AI systems "gamed" their specifications in unexpected ways \cite{Krakovna2018}. Additionally, AI developers have in some cases provided retrospective analyses of particular AI incidents, such as with Microsoft's "Tay" chatbot \cite{Lee2016}.} The pilot was built using publicly available information through a set of volunteers and contractors manually collecting known AI incidents where AI caused harm in the real world. 

Improving the ability and incentive of AI developers to report incidents requires building additional infrastructure, analogous to the infrastructure that exists for reporting incidents in other domains such as cybersecurity. Infrastructure to support incident sharing that involves non-public information would require the following resources:

\begin{itemize}
    \item Transparent and robust processes to protect organizations from undue reputational harm brought about by the publication of previously unshared incidents. This could be achieved by anonymizing incident information to protect the identity of the organization sharing it. Other information-sharing methods should be explored that would mitigate reputational risk to organizations, while preserving the usefulness of information shared;
    \item A trusted neutral third party that works with each organization under a non-disclosure  agreement to collect and anonymize private information;
    \item An organization that maintains and administers an online platform where users can easily access the incident database, including strong encryption and password protection for private incidents as well as a way to submit new information. This organization would not have to be the same as the third party that collects and anonymizes private incident data;
    \item Resources and channels to publicize the existence of this database as a centralized resource, to accelerate both contributions to the database and positive uses of the knowledge from the database; and
    \item Dedicated researchers who monitor incidents in the database in order to identify patterns and shareable lessons.
\end{itemize}

The costs of incident sharing (e.g., public relations risks) are concentrated on the sharing organization, although the benefits are shared broadly by those who gain valuable information about AI incidents. Thus, a cooperative approach needs to be taken for incident sharing that addresses the potential downsides. A more robust infrastructure for incident sharing (as outlined above), including options for anonymized reporting, would help ensure that fear of negative repercussions from sharing does not prevent the benefits of such sharing from being realized.\footnote{We do not mean to claim that building and using such infrastructure would be sufficient to ensure that AI incidents are addressed effectively. Sharing is only one part of the puzzle for effectively managing incidents. For example, attention should also be paid to ways in which organizations developing AI, and particularly safety-critical AI, can become "high reliability organizations" (see, e.g., \cite{Dietterich2018}).}

\textbf{Recommendation: AI developers should share more information about AI incidents, including through collaborative channels.}

Developers should seek to share AI incidents with a broad audience so as to maximize their usefulness, and take advantage of collaborative channels such as centralized incident databases as that infrastructure matures. In addition, they should move towards publicizing their commitment to (and procedures for) doing such sharing in a routine way rather than in an ad-hoc fashion, in order to strengthen these practices as norms within the AI development community. 

Incident sharing is closely related to but distinct from responsible publication practices in AI and coordinated disclosure of cybersecurity vulnerabilities \cite{Ovadya2019}. Beyond implementation of progressively more robust platforms for incident sharing and contributions to such platforms, future work could also explore connections between AI and other domains in more detail, and identify key lessons from other domains in which incident sharing is more mature (such as the nuclear and cybersecurity industries). 

Over the longer term, lessons learned from experimentation and research could crystallize into a mature body of knowledge on different types of AI incidents, reporting processes, and the costs associated with incident sharing. This, in turn, can inform any eventual government efforts to require or incentivize certain forms of incident reporting.

\newpage 
\section{Software Mechanisms and Recommendations}\label{sec:software}

Software mechanisms involve shaping and revealing the functionality of existing AI systems. They can support verification of new types of claims or verify existing claims with higher confidence. This section begins with an overview of the landscape of software mechanisms relevant to verifying claims, and then highlights several key problems, mechanisms, and associated recommendations.

Software mechanisms, like software itself, must be understood in context (with an appreciation for the role of the people involved). Expertise about many software mechanisms is not widespread, which can create challenges for building trust through such mechanisms. For example, an AI developer that wants to provide evidence for the claim that "user data is kept private"  can help build trust in the lab's compliance with a a formal framework such as differential privacy, but non-experts may have in mind a different definition of privacy.\footnote{For example, consider a desideratum for privacy: access to a dataset should not enable an adversary to learn anything about an individual that could not be learned without access to the database. Differential privacy as originally conceived does not guarantee this--rather, it guarantees (to an extent determined by a privacy budget) that one cannot learn whether that individual was in the database in question.} It is thus critical to consider not only which claims can and can't be substantiated with existing mechanisms in theory, but also who is well-positioned to scrutinize these mechanisms in practice.\footnote{In \hyperref[subsec:support]{\textbf{Section 3.3}}, we discuss the role that computing power--in addition to expertise--can play in influencing who can verify which claims.} 

Keeping their limitations in mind, software mechanisms can substantiate claims associated with AI development in various ways that are complementary to institutional and hardware mechanisms. They can allow researchers, auditors, and others to understand the \textbf{internal workings} of any given system. They can also help \textbf{characterize the behavioral profile} of a system over \textbf{a domain of expected usage}. Software mechanisms could support claims such as:
\begin{itemize}
    \item This system is robust to 'natural' distributional shifts \cite{Amodei2016} \cite{Leike2017};
    \item This system is robust even to adversarial examples \cite{Szegedy2014} \cite{Goodfellow2015};
    \item This system has a well-characterized error surface and users have been informed of contexts in which the system would be unsafe to use;
    \item This system's decisions exhibit statistical parity with respect to sensitive demographic attributes\footnote{Conceptions of, and measures for, fairness in machine learning, philosophy, law, and beyond vary widely. See, e.g., Xiang and Raji \cite{Xiang2019} and Binns \cite{Binns2017}.}; and
    \item This system provides repeatable or reproducible results.
\end{itemize}
Below, we summarize several clusters of mechanisms which help to substantiate some of the claims above.

\textbf{Reproducibility} of technical results in AI is a key way of enabling verification of claims about system properties, and a number of ongoing initiatives are aimed at improving reproducibility in AI.\footnote{We note the distinction between narrow senses of reproducibility that focus on discrete technical results being reproducible given the same initial conditions, sometimes referred to as repeatability, and broader senses of reproducibility that involve reported performance gains carrying over to different contexts and implementations.}\footnote{One way to promote robustness is through incentivizing reproducibility of reported results. There are increasing effort to award systems the recognition that they are robust, e.g., through ACM's artifact evaluation badges https://www.acm.org/publications/policies/artifact-review-badging. Conferences are also introducing artifact evaluation, e.g., in the intersection between computer systems research and ML. See, e.g., https://reproindex.com/event/repro-sml2020 and http://cknowledge.org/request.html The Reproducibility Challenge is another notable effort in this area: https://reproducibility-challenge.github.io/neurips2019/} Publication of results, models, and code increase the ability of outside parties (especially technical experts) to verify claims made about AI systems. Careful experimental design and the use of (and contribution to) standard software libraries can also improve reproducibility of particular results.\footnote{In the following section on hardware mechanisms, we also discuss how reproducibility can be advanced in part by leveling the playing field between industry and other sectors with respect to computing power.}

\textbf{Formal verification} establishes whether a system satisfies some requirements using the formal methods of mathematics. Formal verification is often a compulsory technique deployed in various safety-critical domains to provide guarantees regarding the functional behaviors of a system. These are typically guarantees that testing cannot provide. Until recently, AI systems utilizing machine learning (ML)\footnote{Machine learning is a subfield of AI focused on the design of software that improves in response to data, with that data taking the form of unlabeled data, labeled data, or experience. While other forms of AI that do not involve machine learning can still raise privacy concerns, we focus on machine learning here given the recent growth in associated privacy techniques as well as the widespread deployment of machine learning.} have not generally been subjected to such rigor, but the increasing use of ML in safety-critical domains, such as automated transport and robotics, necessitates the creation of novel formal analysis techniques addressing ML models and their accompanying non-ML components. Techniques for formally verifying ML models are still in their infancy and face numerous challenges,\footnote{Research into perception-based properties such as pointwise robustness, for example, are not sufficiently comprehensive to be applied to real-time critical AI systems such as autonomous vehicles.} which we discuss in \hyperref[subsec:analysis]{\textbf{Appendix VI(A)}}.

\textbf{The empirical verification and validation of machine learning by machine learning} has been proposed as an alternative paradigm to formal verification. Notably, it can be more practical than formal verification, but since it operates empirically, the method cannot as fully guarantee its claims. Machine learning could be used to search for common error patterns in another system's code, or be used to create simulation environments to adversarially find faults in an AI system's behavior. 

For example, adaptive stress testing (AST) of an AI system allows users to find the most likely failure of a system for a given scenario using reinforcement learning \cite{Koren2019}, and is being used by to validate the next generation of aircraft collision avoidance software \cite{Lee2015}. Techniques requiring further research include using machine learning to evaluate another machine learning system (either by directly inspecting its policy or by creating environments to test the model) and using ML to evaluate the input of another machine learning model. In the future, data from model failures, especially pooled across multiple labs and stakeholders, could potentially be used to create classifiers that detect suspicious or anomalous AI behavior.

\textbf{Practical verification} is the use of scientific protocols to characterize a model's data, assumptions, and performance. Training data can be rigorously evaluated for representativeness \cite{Chawla2002} \cite{Lemaitre2016}; assumptions can be characterized by evaluating modular components of an AI model and by clearly communicating output uncertainties; and performance can be characterized by measuring generalization, fairness, and performance heterogeneity across population subsets. Causes of differences in performance between models could be robustly attributed via randomized controlled trials.

A developer may wish to make claims about a system's \textbf{adversarial robustness}.\footnote{Adversarial robustness refers to an AI system's ability to perform well in the context of (i.e. to be robust against) "adversarial" inputs, or inputs designed specifically to degrade the system's performance.} Currently, the security balance is tilted in favor of attacks rather than defenses, with only \textit{adversarial training} \cite{Madry2017} having stood the test of multiple years of attack research. \textit{Certificates of robustness}, based on formal proofs, are typically approximate and give meaningful bounds of the increase in error for only a limited range of inputs, and often only around the data available for certification (i.e. not generalizing well to unseen data \cite{Hein2017} \cite{Lecuyer2018} \cite{Cohen2019}). Without approximation, certificates are computationally prohibitive for all but the smallest real world tasks \cite{Katz2017}. Further, research is needed on scaling formal certification methods to larger model sizes.

The subsections below discuss software mechanisms that we consider especially important to advance further. In particular, we discuss \hyperref[subsec:trails]{\textbf{audit trails}}, \hyperref[subsec:interpretabilitysoft]{\textbf{interpretability}}, and \hyperref[subsec:ppml]{\textbf{privacy-preserving machine learning}}.

\newpage 

\subsection{Audit Trails}\label{subsec:trails}
\setlength{\fboxsep}{6pt}
\textbf{Problem:}

\vspace{-0.5\baselineskip}
\fbox{%
    \parbox{0.97\textwidth}{%
        \textbf{AI systems lack traceable logs of steps taken in problem-definition, design, development, and operation, leading to a lack of accountability for subsequent claims about those systems' properties and impacts.}
    }%
}

Audit trails can improve the verifiability of claims about engineered systems, although they are not yet a mature mechanism in the context of AI. An audit trail is a traceable log of steps in system operation, and potentially also in design and testing. We expect that audit trails will grow in importance as AI is applied to more safety-critical contexts. They will be crucial in supporting many institutional trust-building mechanisms, such as third-party auditors, government regulatory bodies,\footnote{Such as the National Transportation Safety Board with regards to autonomous vehicle traffic accidents.} and voluntary disclosure of safety-relevant information by companies.

Audit trails could cover all steps of the AI development process, from the institutional work of problem and purpose definition leading up to the initial creation of a system, to the training and development of that system, all the way to retrospective accident analysis.

There is already strong precedence for audit trails in numerous industries, in particular for safety-critical systems. Commercial aircraft, for example, are equipped with flight data recorders that record and capture multiple types of data each second \cite{Bonsor}. In safety-critical domains, the compliance of such evidence is usually assessed within a larger "assurance case" utilising the CAE or Goal-Structuring-Notation (GSN) frameworks.\footnote{See \hyperref[subsec:importance]{\textbf{Appendix III}} for discussion of assurance cases and related frameworks.} Tools such as the Assurance and Safety Case Environment (ACSE) exist to help both the auditor and the auditee manage compliance claims and corresponding evidence. Version control tools such as GitHub or GitLab can be utilized to demonstrate individual document traceability. Proposed projects like Verifiable Data Audit \cite{Suleyman2017} could establish confidence in logs of data interactions and usage.

\textbf{Recommendation: Standards setting bodies should work with academia and industry to develop audit trail requirements for safety-critical applications of AI systems.}

Organizations involved in setting technical standards--including governments and private actors--should establish clear guidance regarding how to make safety-critical AI systems fully auditable.\footnote{Others have argued for the importance of audit trails for AI elsewhere, sometimes under the banner of "logging." See, e.g., \cite{Bryson2018}.} Although application dependent, software audit trails often require a base set of traceability\footnote{Traceability in this context refers to "the ability to verify the history, location, or application of an item by means of documented recorded identification," https://en.wikipedia.org/wiki/Traceability, where the item in question is digital in nature, and might relate to various aspects of an AI system's development and deployment process.} trails to be demonstrated for qualification;\footnote{This includes traceability: between the system safety requirements and the software safety requirements, between the software safety requirements specification and software architecture, between the software safety requirements specification and software design, between the software design specification and the module and integration test specifications, between the system and software design requirements for hardware/software integration and the hardware/software integration test specifications, between the software safety requirements specification and the software safety validation plan, and between the software design specification and the software verification (including data verification) plan.} the decision to choose a certain set of trails requires considering trade-offs about efficiency, completeness, tamperproofing, and other design considerations. There is flexibility in the type of documents or evidence the auditee presents to satisfy these general traceability requirements (e.g., between test logs and requirement documents, verification and validation activities, etc.).\footnote{See \hyperref[subsec:importance]{\textbf{Appendix III}}.}

Existing standards often define in detail the required audit trails for specific applications. For example, IEC 61508 is a basic functional safety standard required by many industries, including nuclear power. Such standards are not yet established for AI systems. A wide array of audit trails related to an AI development process can already be produced, such as code changes, logs of training runs, all outputs of a model, etc. Inspiration might be taken from recent work on internal algorithmic auditing \cite{Raji2020} and ongoing work on the documentation of AI systems more generally, such as the ABOUT ML project \cite{Raji2019}. Importantly, we recommend that in order to have maximal impact, any standards for AI audit trails should be published freely, rather than requiring payment as is often the case.

\newpage 

\subsection{Interpretability}\label{subsec:interpretabilitysoft}
\setlength{\fboxsep}{6pt}
\textbf{Problem:}

\vspace{-0.5\baselineskip}
\fbox{%
    \parbox{0.97\textwidth}{%
        \textbf{It's difficult to verify claims about "black-box" AI systems that make predictions without explanations or visibility into their inner workings. This problem is compounded by a lack of consensus on what interpretability means.}
    }%
}

Despite remarkable performance on a variety of problems, AI systems are frequently termed "black boxes" due to the perceived difficulty of understanding and anticipating their behavior. This lack of \emph{interpretability} in AI systems has raised concerns about using AI models in high stakes decision-making contexts where human welfare may be compromised \cite{Rudin2018}. Having a better understanding of how the internal processes within these systems work can help proactively anticipate points of failure, audit model behavior, and inspire approaches for new systems. 

Research in model interpretability is aimed at helping to understand how and why a particular model works. A precise, technical definition for interpretability is elusive; by nature, the definition is subject to the inquirer. Characterizing desiderata for interpretable models is a helpful way to formalize interpretability \cite{Lipton2016} \cite{Sundararajan2017}.  Useful interpretability tools for building trust are also highly dependent on the target user and the downstream task. For example, a model developer or regulator may be more interested in understanding model behavior over the entire input distribution whereas a novice layperson may wish to understand why the model made a particular prediction for their individual case.\footnote{While definitions in this area are contested, some would distinguish between "interpretability" and "explainability" as categories for these two directions, respectively.}

Crucially, an "interpretable" model may not be necessary for all situations. The weight we place upon a model being interpretable may depend upon a few different factors, for example:
\begin{itemize}
    \item More emphasis in sensitive domains (e.g., autonomous driving or healthcare,\footnote{See, e.g., Sendak et. al. \cite{Sendak2019} which focuses on building trust in a hospital context, and contextualizes the role of interpretability in this process.} where an incorrect prediction adversely impacts human welfare) or when it is important for end-users to have actionable recourse (e.g., bank loans) \cite{Ustun2019};
    \item Less emphasis given historical performance data (e.g., a model with sufficient historical performance may be used even if it's not interpretable); and
    \item Less emphasis if improving interpretability incurs other costs (e.g., compromising privacy).
\end{itemize}

In the longer term, for sensitive domains where human rights and/or welfare can be harmed, we anticipate that interpretability will be a key component of AI system audits, and that certain applications of AI will be gated on the success of providing adequate intuition to auditors about the model behavior. This is already the case in regulated domains such as finance \cite{Poursabzi-Sangdeh2018}.\footnote{In New York, an investigation is ongoing into apparent gender discrimination associated with the Apple Card's credit line allowances. This case illustrates the interplay of (a lack of) interpretability and the potential harms associated with automated decision-making systems \cite{Vigdor2019}.}

An ascendent topic of research is how to compare the relative merits of different interpretability methods in a sensible way. Two criteria appear to be crucial: \underline{a.} The method should provide sufficient insight for the end-user to understand how the model is making its predictions (e.g., to assess if it aligns with human judgment), and \underline{b.} the interpretable explanation should be faithful to the model, i.e., accurately reflect its underlying behavior.

Work on evaluating \underline{a.}, while limited in treatment, has primarily centered on comparing methods using human surveys \cite{Bau2019}. More work at the intersection of human-computer interaction, cognitive science, and interpretability research--e.g., studying the efficacy of interpretability tools or exploring possible interfaces--would be welcome, as would further exploration of how practitioners currently use such tools \cite{Hohman2019} \cite{Lage2019} \cite{Dietvorst2015} \cite{Poursabzi-Sangdeh2018} \cite{Bhatt2019}.

Evaluating \underline{b.}, the reliability of existing methods is an active area of research \cite{Adebayo2018} \cite{Mahendran2016} \cite{Kindermans2017} \cite{Hooker2018} \cite{Gu2017} \cite{Heo2019} \cite{Slack2019} \cite{Ribeiro2016a} \cite{Dimanov2020}. This effort is complicated by the lack of ground truth on system behavior (if we could reliably anticipate model behavior under all circumstances, we would not need an interpretability method). The wide use of interpretable tools in sensitive domains underscores the continued need to develop benchmarks that assess the reliability of produced model explanations.

It is important that techniques developed under the umbrella of interpretability not be used to provide clear explanations when such clarity is not feasible. Without sufficient rigor, interpretability could be used in service of \textit{unjustified trust} by providing misleading explanations for system behavior. In identifying, carrying out, and/or funding research on interpretability, particular attention should be paid to whether and how such research might eventually aid in verifying claims about AI systems with high degrees of confidence to support risk assessment and auditing.

\textbf{Recommendation: Organizations developing AI and funding bodies should support research into the interpretability of AI systems, with a focus on supporting risk assessment and auditing.}

Some areas of interpretability research are more developed than others. For example, attribution methods for explaining individual predictions of computer vision models are arguably one of the most well-developed research areas. As such, we suggest that the following under-explored directions would be useful for the development of interpretability tools that could support verifiable claims about system properties:

\begin{itemize}
    \item Developing and establishing consensus on the criteria, objectives, and frameworks for interpretability research;
    \item Studying the provenance of a learned model (e.g., as a function of the distribution of training data, choice of particular model families, or optimization) instead of treating models as fixed; and  
    \item Constraining models to be interpretable by default, in contrast to the standard setting of trying to interpret a model post-hoc.
\end{itemize}

This list is not intended to be exhaustive, and we recognize that there is uncertainty about which research directions will ultimately bear fruit. We discuss the landscape of interpretability research further in \hyperref[subsec:interpretability]{\textbf{Appendix VI(C)}}.

\newpage 

\subsection{Privacy-Preserving Machine Learning}\label{subsec:ppml}
\setlength{\fboxsep}{6pt}
\textbf{Problem:}

\vspace{-0.5\baselineskip}
\fbox{%
    \parbox{0.97\textwidth}{%
        \textbf{A range of methods can potentially be used to verifiably safeguard the data and models involved in AI development. However, standards are lacking for evaluating new privacy-preserving machine learning techniques, and the ability to implement them currently lies outside a typical AI developer's skill set.}
    }%
}

Training datasets for AI often include sensitive information about people, raising risks of privacy violation. These risks include unacceptable access to raw data (e.g., in the case of an untrusted employee or a data breach), unacceptable inference from a trained model (e.g., when sensitive private information can be extracted from a model), or unacceptable access to a model itself (e.g., when the model represents personalized preferences of an individual or is protected by intellectual property). 

For individuals to trust claims about an ML system sufficiently so as to participate in its training, they need evidence about data access (who will have access to what kinds of data under what circumstances), data usage, and data protection. The AI development community, and other relevant communities, have developed a range of methods and mechanisms to address these concerns, under the general heading of "privacy-preserving machine learning" (PPML) \cite{Chaudhuri2018}.

Privacy-preserving machine learning aims to protect the privacy of data or models used in machine learning, at training or evaluation time and during deployment. PPML has benefits for model users, and for those who produce the data that models are trained on. 

PPML is heavily inspired by research from the cryptography and privacy communities and is performed in practice using a combination of techniques, each with its own limitations and costs. These techniques are a powerful tool for supporting trust between data owners and model users, by ensuring privacy of key information. However, they must be used judiciously, with informed trade-offs among (1) privacy benefits, (2) model quality, (3) AI developer experience and productivity, and (4) overhead costs such as computation, communication, or energy consumption. They are also not useful in all contexts; therefore, a combination of techniques may be required in some contexts to protect data and models from the actions of well-resourced malicious actors.

Before turning to our recommendation, we provide brief summaries of several PPML techniques that could support verifiable claims. 

\textbf{Federated learning} is a machine learning technique where many clients (e.g., mobile devices or whole organizations) collaboratively train a model under the orchestration of a central server (e.g., service provider), while keeping the training data decentralized \cite{Kairouz2019}. Each client's raw data is stored locally and not exchanged or transferred \cite{Kairouz2019}. Federated learning addresses privacy concerns around the centralized collection of raw data, by keeping the data where it is generated (e.g., on the user's device or in a local silo) and only allowing model updates to leave the client. 

Federated learning does not, however, fully guarantee the privacy of sensitive data on its own, as some aspects of raw data could be memorized in the training process and extracted from the trained model if measures are not taken to address this threat. These measures include quantifying the degree to which models memorize training data \cite{Carlini2018}, and incorporating differential privacy techniques to limit the contribution of individual clients in the federated setting \cite{McMahan2017}. Even when used by itself, federated learning addresses the threats that are endemic to centralized data collection and access, such as unauthorized access, data hacks, and leaks, and the inability of data owners to control their data lifecycle.

\textbf{Differential privacy} \cite{Dwork2006} is a system for publicly sharing information derived from a dataset by describing the patterns of groups within the dataset, while withholding information about individuals in the dataset; it allows for precise measurements of privacy risks for current and potential data owners, and can address the raw-data-extraction threat described above. Differential privacy works through the addition of a controlled amount of statistical noise to obscure the data contributions from records or individuals in the dataset.\footnote{To illustrate how statistical noise can be helpful in protecting privacy, consider the difference between a survey that solicits and retains "raw" answers from individuals, on the one hand, and another survey in which the respondents are asked to flip a coin in order to determine whether they will either provide the honest answer right away or flip the coin again in order to determine which answer to provide. The latter approach would enable individual survey respondents to have plausible deniability regarding their true answers, but those conducting the survey could still learn useful information from the responses, since the noise would largely cancel out at scale. For an accessible discussion of the ideas behind differential privacy and its applications, from which this short summary was adapted, see \cite{Roth2019}.} Differential privacy is already used in various private and public AI settings, and researchers are exploring its role in compliance with new privacy regulations \cite{Cummings2018} \cite{Roth2019}.

Differential privacy and federated learning complement each other in protecting the privacy of raw data: federated learning keeps the raw data on the personal device, so it is never seen by the model trainer, while differential privacy ensures the model sufficiently prevents the memorization of  raw data, so that it cannot be extracted from the model by its users.\footnote{For an example of the combination of federated learning and differential privacy, see McMahan et. al. \cite{McMahan2017}.} These techniques do not, however, protect the model itself from theft \cite{Isakov2019}.

\textbf{Encrypted computation} addresses this risk by allowing the model to train and run on encrypted data while in an encrypted state, at the cost of overhead in terms of computation and communication. As a result, those training the model will not be able to see, leak, or otherwise abuse the data in its unencrypted form. The most well known methods for encrypted computation are homomorphic encryption, secure multi-party computation, and functional encryption \cite{Dowlin2016}. For example, one of OpenMined's upcoming projects is Encrypted Machine Learning as a Service, which allows a model owner and data owner to use their model and data to make a prediction, without the model owner disclosing their model, and without the data owner disclosing their data.\footnote{See \url{https://www.openmined.com}.} 

These software mechanisms can guarantee tighter bounds on AI model usage than the legal agreements that developers currently employ, and tighter bounds on user data usage than institutional mechanisms such as user privacy agreements. Encrypted computation could also potentially improve the verifiability of claims by allowing sensitive models to be shared for auditing in a more secure fashion. A hardware-based method to protect models from theft (and help protect raw data from adversaries) is the use of secure enclaves, as discussed in \hyperref[subsec:securehardware]{\textbf{Section 4.1}} below.

In the future, it may be possible to rely on a platform that enables \textbf{verifiable data policies} which address some of the security and privacy vulnerabilities in existing IT systems. One proposal for such a platform is Google's Project Oak,\footnote{See the \hyperref[subsec:distributed]{\textbf{Appendix VI(B)}} for further discussion of this project.} which leverages open source secure enclaves (see \hyperref[subsec:securehardware]{\textbf{Section 4.1}}) and formal verification to technically enforce and assure policies around data storage, manipulation, and exchange.

As suggested by this brief overview of PPML techniques, there are many opportunities for improving the privacy and security protections associated with ML systems. However, greater standardization of of PPML techniques--and in particular, the use of open source PPML frameworks that are benchmarked against common performance measures--may be needed in order for this to translate into a major impact on the verifiability of claims about AI development. First, robust open source frameworks are needed in order to reduce the skill requirement for implementing PPML techniques, which to date have primarily been adopted by large technology companies with in-house expertise in both ML and cryptography. Second, common standards for evaluating new PPML techniques could increase the comparability of new results, potentially accelerating research progress. Finally, standardization could improve the ability of external parties (including users, auditors, and policymakers) to verify claims about PPML performance. 

\textbf{Recommendation: AI developers should develop, share, and use suites of tools for privacy-\\preserving machine learning that include measures of performance against common standards.}

Where possible, AI developers should contribute to, use, and otherwise support the work of open-source communities working on PPML, such as OpenMined, Microsoft SEAL, tf-encrypted, tf-federated, and nGraph-HE. These communities have opened up the ability to use security and privacy tools in the ML setting, and further maturation of the software libraries built by these communities could yield still further benefits. 

Open-source communities projects or projects backed by a particular company can sometimes suffer from a lack of stable funding support\footnote{Novel approaches to funding open source work should also be considered in this context, such as GitHub's "sponsors" initiative. \url{https://help.github.com/en/github/supporting-the-open-source-community-with-github-sponsors/about-github-sponsors}} or independence as organizational priorities shift, suggesting a need for an AI community-wide approach to supporting PPML's growth. Notwithstanding some challenges associated with open source projects, they are uniquely amenable to broad-based scrutiny and iteration, and have yielded benefits already. Notably, integrated libraries for multiple techniques in privacy-preserving ML have started being developed for major deep learning frameworks such as TensorFlow and PyTorch.

Benchmarks for PPML could help unify goals and measure progress across different groups.\footnote{The use of standard tools, guides, and benchmarks can also potentially advance research in other areas, but we focus on privacy-preserving ML in particular here given the backgrounds of the authors who contributed to this subsection. Additionally, we note that some benchmarks have been proposed in the PPML literature for specific subsets of techniques, such as DPComp for differential privacy, but we expect that further exploration of benchmarks across the full spectra of PPML techniques would be valuable.} A centralized repository of real-world implementation cases, a compilation of implementation guides, and work on standardization/interoperability would all also aid in supporting adoption and scrutiny of privacy-preserving methods.\footnote{On the other hand, we note that benchmarks also have potential disadvantages, as they incentivize developers to perform well on the specific benchmark, rather than focusing on the specifics of the intended use case of their product or service, which may significantly diverge from the benchmark setting; the design of benchmarks, and more diverse and adaptive evaluation and comparison methods, is its own technical challenge, as well as an institutional challenge to incentivize appropriate curation and use of benchmarks to establish a common understanding of what is achievable.}

\newpage 
\section{Hardware Mechanisms and Recommendations}\label{sec:hardware}

Computing hardware enables the training, testing, and use of AI systems. Hardware relevant to AI development ranges from sensors, networking, and memory, to, perhaps most crucially, processing power \cite{Hwang2018}.\footnote{Experts disagree on the extent to which large amounts of computing power are key to progress in AI development. See, e.g., Sutton \cite{Sutton2019} and Brooks \cite{Brooks2019} for different opinions about the importance of computing power relative to other factors.} Concerns about the security and other properties of computing hardware, as well as methods to address those concerns in a verifiable manner, long precede the current growth in adoption of AI. However, because of the increasing capabilities and impacts of AI systems and the particular hardware demands of the field, there is a need for novel approaches to assuring the verifiability of claims about the hardware used in AI development.

Hardware mechanisms involve physical computing resources (e.g., CPUs and GPUs), including their distribution across actors, the ways they are accessed and monitored, and their properties (e.g., how they are designed, manufactured, or tested). Hardware can support verifiable claims in various ways. Secure hardware can play a key role in private and secure machine learning by translating privacy constraints and security guarantees into scrutable hardware designs or by leveraging hardware components in a software mechanism. Hardware mechanisms can also be used to demonstrate the ways in which an organization is using its general-purpose computing capabilities. 

At a higher level, the distribution of computing power across actors can potentially influence who is in a position to verify certain claims about AI development. This is true on the assumption that, all things being equal, more computing power will enable more powerful AI systems to be built, and that a technical capability to verify claims may itself require non-negligible computing resources.\footnote{Since training AI systems is more compute-intensive\footnote{"Compute" is commonly used as a more concise way of referring to computing power.} than running them, it is not clear that equivalent computational resources will always be required on the part of those verifying claims about an AI system. However, AI systems are also beginning to require non-trivial computing resources to run, sometimes requiring the model to be split over multiple machines. Additionally, one might need to run an AI system many times in order to verify claims about its characteristics, even if each run is inexpensive. We thus make the conservative assumption that more computing resources would be (all things being equal) helpful to the scrutiny of claims about large-scale AI systems, as discussed below in the context of academic access to computing resources, while recognizing that this may not always be true in particular cases.} The use of standardized, publicly available hardware (sometimes called "commodity hardware") across AI systems also aids in the independent reproducibility of technical results, which in turn could play a role in technical auditing and other forms of accountability. Finally, hardware mechanisms can be deployed to  enforce and verify policies relating to the security of the hardware itself (which, like software, might be compromised through error or malice). 

Existing mechanisms performing one or more of these functions are discussed below.

\textbf{Formal verification}, discussed above in the software mechanisms section, is the process of establishing whether a software or hardware system satisfies some requirements or properties, using formal methods to generate mathematical proofs. Practical tools, such as GPUVerify for GPU kernels,\footnote{http://multicore.doc.ic.ac.uk/tools/GPUVerify/} exist to formally verify components of the AI hardware base, but verification of the complete hardware base is currently an ambitious goal. Because only parts of the AI hardware ecosystem are verified, it is important to map which properties are being verified for different AI accelerators and under what assumptions, who has access to evidence of such verification processes (which may be part of a third party audit), and what properties we should invest more research effort into verifying (or which assumption would be a priority to drop).

\textbf{Remote attestation} leverages a "root of trust" (provided in hardware or in software, e.g., a secret key stored in isolated memory) to cryptographically sign a measurement or property of the system, thus providing a remote party proof of the authenticity of the measurement or property. Remote attestation is often used to attest that a certain version of software is currently running, or that a computation took a certain amount of time (which can then be compared to a reference by the remote party to detect tampering) \cite{Chen2019}.

\textbf{Cloud computing:} Hardware is also at the heart of the relationship between cloud providers and cloud users (as hardware resources are being rented). Associated verification mechanisms can help ensure that computations are being performed as promised, without the client having direct physical access to the hardware. For example, one could have assurances that a cloud-based AI service is not skimping on computations by running a less powerful model than advertised, operating on private data in a disallowed fashion, or compromised by malware \cite{Ghodsi2017}. 

Cloud providers are a promising intervention point for trust-building mechanisms; a single cloud provider services, and therefore has influence over, many customers. Even large AI labs rely predominantly on cloud computing for some or all of their AI development. Cloud providers already employ a variety of mechanisms to minimize risks of misuse on their platforms, including "Know Your Customer" services and Acceptable Use Policies. These mechanisms could be extended to cover AI misuse \cite{Javadi2020}. Additional mechanisms could be developed such as a forum where cloud providers can share best-practices about detecting and responding to misuse and abuse of AI through their services.\footnote{These conversations could take place in existing industry fora, such as the Cloud Security Alliance (https://cloudsecurityalliance.org), or through the establishment of new fora dedicated to AI cloud providers.}

We now turn to more detailed discussions of three hardware mechanisms that could improve the verifiability of claims: we highlight the importance of \hyperref[subsec:securehardware]{\textbf{secure hardware for machine learning}}, \hyperref[subsec:measurement]{\textbf{high-precision compute measurement}}, and \hyperref[subsec:support]{\textbf{computing power support for academia}}.

\newpage 

\subsection{Secure Hardware for Machine Learning}\label{subsec:securehardware}
\setlength{\fboxsep}{6pt}
\textbf{Problem:}

\vspace{-0.5\baselineskip}
\fbox{%
    \parbox{0.97\textwidth}{%
        \textbf{Hardware security features can provide strong assurances against theft of data and models, but secure enclaves (also known as Trusted Execution Environments) are only available on commodity (non-specialized) hardware. Machine learning tasks are increasingly executed on specialized hardware accelerators, for which the development of secure enclaves faces significant up-front costs and may not be the most appropriate hardware-based solution.}
    }%
}

Since AI systems always involve physical infrastructure, the security of that infrastructure can play a key role in claims about a system or its components being secure and private. Secure enclaves have emerged in recent years as a way to demonstrate strong claims about privacy and security that cannot be achieved through software alone. iPhones equipped with facial recognition for screen unlocking, for example, store face-related data on a physically distinct part of the computer known as a secure enclave in order to provide more robust privacy protection. Increasing the range of scenarios in which secure enclaves can be applied in AI, as discussed in this subsection, would enable higher degrees of security and privacy protection to be demonstrated and demanded. 

A secure enclave is a set of software and hardware features that together provide an isolated execution environment that enables a set of strong guarantees regarding security for applications running inside the enclave \cite{Mckeen2013}. Secure enclaves reduce the ability of malicious actors to access sensitive data or interfere with a program, even if they have managed to gain access to the system outside the enclave. Secure enclaves provide these guarantees by linking high-level desired properties (e.g., isolation of a process from the rest of the system) to low-level design of the chip layout and low-level software interacting with the chip. 

The connection between physical design and low-level software and high-level security claims relies on a set of underlying assumptions. Despite the fact that researchers have been able to find ways to invalidate these underlying assumptions in some cases, and thus invalidate the high-level security claims \cite{VanBulck2018} \cite{Seaborn2015},  these mechanisms help to focus defensive efforts and assure users that relatively extreme measures would be required to invalidate the claims guaranteed by the design of the enclave.

While use of secure enclaves has become relatively commonplace in the commodity computing industries, their use in machine learning is less mature. Execution of machine learning on secure enclaves has been demonstrated, but comes with a performance overhead \cite{Hynes2018}.\footnote{For example, training ResNet-32 using Myelin (which utilises a CPU secure enclave) requires 12.9 mins/epoch and results in a final accuracy of 90.8\%, whereas the same training on a non-private CPU requires 12.3 mins/epoch and results in a final accuracy of 92.4\% \cite{Hynes2018}.} Demonstrations to date have been carried out on commodity hardware (CPUs \cite{Priebe2019} \cite{Hesamifard2018}  and GPUs \cite{Volos2018}) or have secure and verifiable outsourcing of parts of the computation to less secure hardware \cite{Tramer2018} \cite{Gu2018},  rather than on hardware directly optimized for machine learning (such as TPUs).

For most machine learning applications, the cost of using commodity hardware not specialized for machine learning is fairly low because the hardware already exists, and their computational demands can be met on such commodity hardware. However, cutting edge machine learning models often use significantly more computational resources \cite{Amodei2018}, driving the use of more specialized hardware for both training and inference. If used with specialized AI hardware, the use of secure enclaves would require renewed investment for every new design, which can end up being very costly if generations are short and of limited batch sizes (as the cost is amortized across all chips that use the design). Some specialized AI hardware layouts may require entirely novel hardware security features -- as the secure enclave model may not be applicable -- involving additional costs. 

One particularly promising guarantee that might be provided by ML-specific hardware security features, coupled with some form of remote attestation, is a guarantee that a model will never leave a particular chip, which could be a key building block of more complex privacy and security policies.

\textbf{Recommendation: Industry and academia should work together to develop hardware security features for AI accelerators\footnote{An AI accelerator is a form of computing hardware that is specialized to perform an AI-related computation efficiently, rather than to perform general purpose computation.} or otherwise establish best practices for the use of secure hardware (including secure enclaves on commodity hardware) in machine learning contexts.}

A focused and ongoing effort to integrate hardware security features into ML-specialized hardware could add value, though it will require collaboration across the sector. 

Recent efforts to open source secure enclave designs could help accelerate the process of comprehensively analyzing the security claims made about certain systems \cite{Newman2019}. As more workloads move to specialized hardware, it will be important to either develop secure enclaves for such hardware (or alternative hardware security solutions), or otherwise define best practices for outsourcing computation to "untrusted" accelerators while maintaining privacy and security. Similarly, as many machine learning applications are run on GPUs, it will be important to improve the practicality of secure enclaves or equivalent privacy protections on these processors.

The addition of dedicated security features to ML accelerators at the hardware level may need to take a different form than a secure enclave. This is in part due to different architectures and different use of space on the chip; in part due to different weighting of security concerns (e.g., it may be especially important to prevent unauthorized access to user data); and in part due to a difference in economies of scale relative to commodity chips, with many developers of ML accelerators being smaller, less-well-resourced actors relative to established chip design companies like Intel or NVIDIA.

\newpage 

\subsection{High-Precision Compute Measurement}\label{subsec:measurement}
\setlength{\fboxsep}{6pt}
\textbf{Problem:}

\vspace{-0.5\baselineskip}
\fbox{%
    \parbox{0.97\textwidth}{%
        \textbf{The absence of standards for measuring the use of computational resources reduces the value of voluntary reporting and makes it harder to verify claims about the resources used in the AI development process.}
    }%
}

Although we cannot know for certain due to limited transparency, it is reasonable to assume that a significant majority of contemporary computing hardware used for AI training and inference is installed in data centers (which could be corporate, governmental, or academic), with smaller fractions in server rooms or attached to individual PCs.\footnote{For reference, the Cisco Global Cloud Index forecasts that the ratio of data center traffic to non-data center traffic by 2021 will be 103:1. When looking just at data centers, they forecast that by 2021, 94\% of workloads and compute instances will be processed by cloud data centers, with the remaining 6\% processed by traditional data centers. Note, however, that these are for general workloads, not AI specific \cite{Cisco2018}.} 

Many tools and systems already exist to monitor installed hardware and compute usage internally (e.g., across a cloud provider's data center or across an academic cluster's user base). A current example of AI developers reporting on their compute usage is the inclusion of training-related details in published research papers and pre-prints, which often share the amount of compute used to train or run a model.\footnote{Schwartz and Dodge et al. have recommended that researchers always publish financial and computational costs alongside performance increases \cite{Schwartz2019}.} These are done for the purposes of comparison and replication, though often extra work is required to make direct comparisons as there is no standard method for measuring and reporting compute usage.\footnote{There are, however, emerging efforts at standardization in particular contexts. The Transaction Processing Performance Council has a related working group, and efforts like MLPerf are contributing to standardization of some inference-related calculations, though accounting for training remains especially problematic.} This ambiguity poses a challenge to trustworthy AI development, since even AI developers who want to make verifiable claims about their hardware use are not able to provide such information in a standard form that is comparable across organizations and contexts.

Even in the context of a particular research project, issues such as mixed precision training,\footnote{Mixed precision refers to the growing use of different binary representations of floating point numbers with varying levels of precision (e.g., 8 bit, 16 bit, 32 bit or 64 bit) at different points of a computation, often trading-off lower precision for higher throughput or performance in ML contexts relative to non-ML contexts. Since an 8 bit floating point operation, say,  differs in hardware requirements from a 64 bit floating point operation, traditional measures in terms of Floating Point Operations Per Second (FLOPS) fail to capture this heterogeneity.} use of heterogeneous computing resources, and use of pretrained models all complicate accurate reporting that is comparable across organizations.\footnote{For an illustrative discussion of the challenges associated with reporting compute usage for a large-scale AI project, see, e.g., OpenAI's Dota 2 project \cite{Berner2019}.} The lack of a common standard or accepted practice on how to report the compute resources used in the context of a particular project has led to several efforts to extract or infer the computational requirements of various advances and compare them using a common framework \cite{Amodei2018}.

The challenge of providing accurate and useful information about the computational requirements of a system or research project is not unique to AI -- computer systems research has struggled with this problem for some time.\footnote{See, e.g., Vitek \& Kalibera \cite{Vitek2011} and Hoefler \& Belli \cite{Hoefler2015}.} Both fields have seen an increasing challenge in comparing and reproducing results now that organizations with exceptionally large compute resources (also referred to as "hyperscalers") play an ever-increasing role in research in those fields. We believe there is value in further engaging with the computer systems research community to explore challenges of reproducibility, benchmarking, and reporting, though we also see value in developing AI-specific standards for compute reporting.

Increasing the precision and standardization of compute reporting could enable easier comparison of research results across organizations. Improved methods could also serve as building blocks of credible third party oversight of AI projects: an auditor might note, for example, that an organization has more computing power available to it than was reportedly used on an audited project, and thereby surface unreported activities relevant to that project. And employees of an organization are better able to ensure that their organization is acting responsibly to the extent that they are aware of how computing power, data, and personnel are being allocated internally for different purposes. 

\textbf{Recommendation: One or more AI labs should estimate the computing power involved in a single project in great detail, and report on the potential for wider adoption of such methods.}

We see value in one or more AI labs conducting a "comprehensive" compute accounting effort, as a means of assessing the feasibility of standardizing such accounting. "Comprehensive" here refers to accounting for as much compute usage pertinent to the project as is feasible, and increasing the precision of reported results relative to existing work.

It is not clear how viable standardization is, given the aforementioned challenges, though there is likely room for at least incremental progress: just in the past few years, a number of approaches to calculating and reporting compute usage have been tried, and in some cases have propagated across organizations. AI researchers interested in conducting such a pilot should work with computer systems researchers who have worked on related challenges in other contexts, including the automating of logging and reporting. 

Notably, accounting of this sort has costs associated with it, and the metrics of success are unclear. Some accounting efforts could be useful for experts but inaccessible to non-experts, for example, or could only be workable in a particular context (e.g., with a relatively simple training and inference pipeline and limited use of pretrained models). As such, we do not advocate for requiring uniformly comprehensive compute reporting. 

Depending on the results of early pilots, new tools might help automate or simplify such reporting, though this is uncertain. One reason for optimism about the development of a standardized approach is that a growing fraction of computing power usage occurs in the cloud at "hyperscale" data centers, so a relatively small number of actors could potentially implement best practices that apply to a large fraction of AI development \cite{Cisco2018}.

It is also at present unclear who should have access to reports about compute accounting. While we applaud the current norm in AI research to voluntarily share compute requirements publicly, we expect for-profit entities would have to balance openness with commercial secrecy, and government labs may need to balance openness with security considerations. This may be another instance in which auditors or independent verifiers could play a role. Standardization of compute accounting is one path to formalizing the auditing practice in this space, potentially as a building block to more holistic auditing regimes. However, as with other mechanisms discussed here, it is insufficient on its own.

\newpage 

\subsection{Compute Support for Academia}\label{subsec:support}
\setlength{\fboxsep}{6pt}
\textbf{Problem:}

\vspace{-0.5\baselineskip}
\fbox{%
    \parbox{0.97\textwidth}{%
        \textbf{The gap in compute resources between industry and academia limits the ability of those outside of industry to scrutinize technical claims made by AI developers, particularly those related to compute-intensive systems.}
    }%
}

In recent years, a large number of academic AI researchers have transitioned into industry AI labs. One reason for this shift is the greater availability of computing resources in industry compared to academia. This talent shift has resulted in a range of widely useful software frameworks and algorithmic insights, but has also raised concerns about the growing disparity between the computational resources available to academia and industry \cite{Lohr2019}.

The disparity between industry and academia is clear overall, even though some academic labs are generously supported by government\footnote{https://epsrc.ukri.org/research/facilities/hpc/} or industry\footnote{See, e.g., https://www.tensorflow.org/tfrc, https://aws.amazon.com/grants/ and https://www.microsoft.com/en-us/research/academic-program/microsoft-azure-for-research/ for examples of industry support for academic computing.} sponsors, and some government agencies are on the cutting edge of building and providing access to supercomputers.\footnote{For example, the US Department of Energy's supercomputing division currently hosts the fastest supercomputer worldwide.}

Here we focus on a specific benefit of governments\footnote{While industry actors can and do provide computing power support to non-industry actors in beneficial ways, the scale and other properties of such programs are likely to be affected by the rises and falls of particular companies' commercial fortunes, and thus are not a reliable long-term solution to the issues discussed here.} taking action to level the playing field of computing power: namely, improving the ability of financially disinterested parties such as academics to verify the claims made by AI developers in industry, especially in the context of compute-intensive systems. Example use cases include:

\begin{itemize}
    \item \textbf{Providing open-source alternatives to commercial AI systems}: given the current norm in AI development of largely-open publication of research, a limiting factor in providing open source alternatives to commercially trained AI models is often the computing resources required. As models become more compute-intensive, government support may be required to maintain a thriving open source AI ecosystem and the various benefits that accrue from it.
    \item \textbf{Increasing scrutiny of commercial models}: as outlined in the institutional mechanisms section (see the subsections on \hyperref[subsec:redteam]{\textbf{red team exercises}} and \hyperref[subsec:bounties]{\textbf{bias and safety bounties}}), there is considerable value in independent third parties stress-testing the models developed by others. While "black box" testing can take place without access to significant compute resources (e.g., by remote access to an instance of the system), local replication for the purpose of testing could make testing easier, and could uncover further issues than those surfaced via remote testing alone. Additional computing resources may be especially needed for local testing of AI systems that are too large to run on a single computer (such as some recent language models).
    \item \textbf{Leveraging AI to test AI}: as AI systems become more complex, it may be useful or even necessary to deploy adaptive, automated tests to explore potential failure modes or hidden biases, and such testing may become increasingly compute-intensive.
    \item \textbf{Verifying claims about compute requirements}: as described above, accounting for the compute inputs of model training is currently an open challenge in AI development. In tandem with standardization of compute accounting, compute support to non-industry actors would enable replication efforts, which would verify or undermine claims made by AI developers about the resource requirements of the systems they develop.
\end{itemize}

\textbf{Recommendation: Government funding bodies should substantially increase funding of computing power resources for researchers in academia, in order to improve the ability of those researchers to verify claims made by industry.}

While computing power is not a panacea for addressing the gap in resources available for research in academia and industry, funding bodies such as those in governments could level the playing field between sectors by more generously providing computing credits to researchers in academia.\footnote{As advocated by various authors, e.g., Sastry et al. \cite{Amodei2018}, Rasser \& Lambert et. al. \cite{Rasser2019}, and Etchemendy and Li \cite{Etchemendy2020}.} Such compute provision could be made more affordable by governments leveraging their purchasing power in negotiations over bulk compute purchases. Governments could also build their own compute infrastructures for this purpose. The particular amounts of compute in question, securing the benefits of scale while avoiding excessive dependence on a particular compute provider, and ways of establishing appropriate terms for the use of such compute are all exciting areas for future research.\footnote{This may require significant levels of funding, and so the benefits should be balanced against the opportunity cost of public spending.}

\newpage 
\section{Conclusion}\label{sec:conclusion}

Artificial intelligence has the potential to transform society in ways both beneficial and harmful. Beneficial applications are more likely to be realized, and risks more likely to be avoided, if AI developers earn rather than assume the trust of society and of one another. This report has fleshed out one way of earning such trust, namely the making and assessment of verifiable claims about AI development through a variety of mechanisms. A richer toolbox of mechanisms for this purpose can inform developers' efforts to earn trust, the demands made of AI developers by activists and civil society organizations, and regulators' efforts to ensure that AI is developed responsibly. 

If the widespread articulation of ethical principles can be seen as a first step toward ensuring responsible AI development, insofar as it helped to establish a standard against which behavior can be judged, then the adoption of mechanisms to make verifiable claims represents a second. The authors of this report are eager to see further steps forward and hope that the framing of these mechanisms inspires the AI community to begin a meaningful dialogue around approaching verifiability in a collaborative fashion across organizations. We are keen to discover, study, and foreground additional institutional, software, and hardware mechanisms that could help enable trustworthy AI development. We encourage readers interested in collaborating in these or other areas to contact the corresponding authors of the report.\footnote{The landing page for this report, www.towardtrustworthyai.com, will also be used to share relevant updates after the report's publication.} 

As suggested by the title of the report (which references supporting verifiable claims rather than ensuring them), we see the mechanisms discussed here as enabling incremental improvements rather than providing a decisive solution to the challenge of verifying claims in the AI ecosystem. And despite the benefits associated with verifiable claims, they are also insufficient to ensure that AI developers will behave responsibly. There are at least three reasons for this. 

First, there is a tension between verifiability of claims and the generality of such claims. This tension arises because the narrow properties of a system are easier to verify than the general ones, which tend to be of greater social interest. Safety writ large, for example, is inherently harder to verify than performance on a particular metric for safety. Additionally, broad claims about the beneficial societal impacts of a system or organization are harder to verify than more circumscribed claims about impacts in specific contexts.     

Second, the verifiability of claims does not ensure that they will be verified in practice. The mere existence of mechanisms for supporting verifiable claims does not ensure that they will be demanded by consumers, citizens, and policymakers (and even if they are, the burden ought not to be on them to do so). For example, consumers often use technologies in ways that are inconsistent with their stated values (e.g., a concern for personal privacy) because other factors such as convenience and brand loyalty also play a role in influencing their behavior \cite{Barth2017}.

Third, even if a claim about AI development is shown to be false, asymmetries of power may prevent corrective steps from being taken. Members of marginalized communities, who often bear the brunt of harms associated with AI \cite{Whittaker2018}, often lack the political power to resist technologies that they deem detrimental to their interests. Regulation will be required to ensure that AI developers provide evidence that bears on important claims they make, to limit applications of AI where there is insufficient technical and social infrastructure for ensuring responsible development, or to increase the variety of viable options available to consumers that are consistent with their stated values. 

These limitations notwithstanding, verifiable claims represent a step toward a more trustworthy AI development ecosystem. Without a collaborative effort between AI developers and other stakeholders to improve the verifiability of claims, society's concerns about AI development are likely to grow: AI is being applied to an increasing range of high-stakes tasks, and with this wide deployment comes a growing range of risks. With a concerted effort to enable verifiable claims about AI development, there is a greater opportunity to positively shape AI's impact and increase the likelihood of widespread societal benefits.

\newpage 
\section*{Acknowledgements}
\phantomsection\addcontentsline{toc}{section}{Acknowledgements}

We are extremely grateful to participants in the April 2019 workshop that catalyzed this report, as well as the following individuals who provided valuable input on earlier versions of this document: David Lansky, Tonii Leach, Shin Shin Hua, Chris Olah, Alexa Hagerty, Madeleine Clare Elish, Larissa Schiavo, Heather Roff,  Rumman Chowdhury, Ludwig Schubert, Joshua Achiam, Chip Huyen, Xiaowei Huang, Rohin Shah, Genevieve Fried, Paul Christiano, Sean McGregor, Tom Arnold, Jess Whittlestone, Irene Solaiman, Ashley Pilipiszyn, Catherine Olsson, Bharath Ramsundar, Brandon Perry, Justin Wang, Max Daniel, Ryan Lowe, Rebecca Crootof, Umang Bhatt, Ben Garfinkel, Claire Leibowicz, Ryan Khurana, Connor Leahy, Chris Berner, Daniela Amodei, Erol Can Akbaba, William Isaac, Iason Gabriel, Laura Weidinger, Thomas Dietterich, Olexa Bilaniuk, and attendees of a seminar talk given by author Miles Brundage on this topic at the Center for Human-Compatible AI (CHAI). None of these people necessarily endorses the content of the report.

\newpage

\phantomsection\addcontentsline{toc}{section}{References}
\printbibliography

\newpage 
\appendix
\section*{Appendices}\label{sec:appendices}
\phantomsection\addcontentsline{toc}{section}{Appendices}
\renewcommand{\thesubsubsection}{\Alph{subsubsection}}
\renewcommand{\thesubsection}{\Roman{subsection}}


\subsection{Workshop and Report Writing Process}\label{subsec:process}

This report began as an effort to identify areas for productive work related to trust in AI development. The project began in earnest with an interdisciplinary expert workshop in San Francisco in April of 2019, which brought together participants from academia, industry labs, and civil society organizations.\footnote{Our intent with this section of the report is to be transparent with readers about our process and to acknowledge some of the voices and methods missing from that process. We also hope that providing information about our process could be helpful for those considering similar multi-stakeholder research projects.} As discussed below, during the writing process, we shifted our focus to verifiable claims in particular, rather than trust more broadly. 

Workshop attendees are listed below in alphabetical order:
\begin{multicols*}{2}
\begin{itemize}
    \item Amanda Askell
    \item Andrew Critch
    \item Andrew Lohn
    \item Andrew Reddie
    \item Andrew Trask
    \item Ben Garfinkel
    \item Brian Tse
    \item Catherine Olsson
    \item Charina Chou
    \item Chris Olah
    \item David Luan
    \item Dawn Song
    \item Emily Oehlsen
    \item Eric Sigler
    \item Genevieve Fried
    \item Gillian Hadfield
    \item Heidy Khlaaf
    \item Helen Toner
\end{itemize}
\vfill\null
\columnbreak
\begin{itemize}
    \item Ivan Vendrov
    \item Jack Clark
    \item Jeff Alstott
    \item Jeremy Nixon
    \item Jingying Yang
    \item Joshua Kroll
    \item Lisa Dyer
    \item Miles Brundage
    \item Molly Welch
    \item Paul Christiano
    \item Peter Eckersley
    \item Se\'{a}n \'{O} h\'{E}igeartaigh
    \item Shahar Avin
    \item Shixiong (Austin) Zhang
    \item Teddy Collins
    \item Tim Hwang
    \item William Isaac
\end{itemize}
\end{multicols*}

Given our initial focus on synthesizing and extending existing work, we brought together experts in dimensions of trust that were identified in a pre-workshop white paper, several of which are discussed in this report (such as secure enclaves, third party auditing, and privacy-preserving machine learning). However, a number of voices were missing from that conversation. The workshop could have benefited in particular from greater gender diversity (fewer than one third of participants were women, and none were members of trans or non-binary communities); greater racial diversity (people of color and especially women of color were under-represented, particularly given the number of women of color with relevant expertise on trust in AI development); greater representation of low income communities; greater representation of people with disabilities; and greater geographic, political, philosophical, and religious diversity. 

Following the workshop, the corresponding authors led a multi-stakeholder writing, editing, and feedback process. A subset of workshop attendees opted to engage in the writing process for the report. After the first round of writing and editing, we tried to incorporate new authors with complementary expertise to those at the original workshop. Notwithstanding these efforts, not all dimensions of trust in AI development (or even verifiable claims in AI development) were represented in the expertise of the authors. As such, over time and especially in response to external reviewer feedback, we progressively narrowed the scope of the report in order to avoid overreach and "stay in our lane" topic-wise. One such shift was a move from discussing trust in AI development generally to verifiable claims specifically as the focus of the report. 

The report was written in a semi-modular fashion. Experts in particular areas drafted subsections on mechanisms or research areas with which they are familiar. These subsections were revised substantially over time in response to author and reviewer feedback, and to shifts in the framing of the report. External reviewers provided feedback on specific portions of the report for clarity and accuracy, although the report as a whole was not formally peer reviewed. 

Given the number of authors involved and the wide-ranging nature of the report, it was difficult to ensure that all authors were fully supportive of all content throughout the writing process. Where appropriate, we have included footnotes to clarify process and attribution.

\newpage

\subsection{Key Terms and Concepts}\label{subsec:terms}

\textbf{AI}: we define artificial intelligence (AI) as any digital system capable of performing tasks commonly thought to require intelligence, with these tasks often being learned from data and/or experience.\footnote{Some distinctions are made between different phases of AI development in the report, although the authors have also found it helpful to take a broad view of look at AI development: in many cases, the same mechanisms (especially institutional ones) are applicable to multiple phases of development, and \textit{AI development} was found to be the most appropriate catch-all term for such purposes. A recent and representative example of a more granular breakdown, from Toreini et al., distinguishes data-related steps (data collection, data preparation, and feature extraction) from model-related steps (training, testing, and inference) \cite{Toreini2019}. In practice, these steps are not followed in a linear manner, since (e.g.) testing may inform changes to the training process and data may be improved over time.}

\textbf{AI system}: we define an AI system as a \textit{software} process (with the characteristics of AI mentioned above), running on physical \textit{hardware}, under the direction of humans operating in some \textit{institutional} context. This framing of AI systems informs the discussion of mechanisms in the report. The properties of the software, hardware, and institutions at work in a given AI system are all potentially relevant to the verifiability of claims made by an AI developer. Focusing on any of these to the exclusion of others could result in a flawed understanding of the overall system.

\textbf{AI development}: we use the term AI development to refer to the process of researching, designing, testing, deploying, or monitoring AI as defined above. 

\textbf{AI developer}: we use the term AI developer to refer to individuals or organizations involved in AI development as defined broadly above, including research scientists, data engineers, and project managers at companies building AI-based products and services as well as those in analogous roles in academia, government, or civil society. Given the major role played by technology companies in contemporary AI development, we pay particular attention to such companies in the report, while recognizing that different contexts will require different approaches to earning trust.

\textbf{Responsible AI development}: we follow Askell et al. in defining responsible AI development as follows \cite{Askell2019}:
\begin{displayquote}
"Responsible AI development involves taking steps to ensure that AI systems have an acceptably low risk of harming their users or society and, ideally, to increase their likelihood of being socially beneficial. This involves testing the safety and security of systems during development, evaluating the potential social impact of the systems prior to release, being willing to abandon research projects that fail to meet a high bar of safety, and being willing to delay the release of a system until it has been established that it does not pose a risk to consumers or the public."
\end{displayquote}

\textbf{Transparency}: we define transparency as making information about the characteristics of an AI developer's operations or their AI systems available to actors both inside and outside the organization. In recent years, transparency has emerged as a key theme in work on the societal implications of AI.\footnote{See, for example, the Fairness, Accountability, and Transparency in Machine Learning (FATML) workshop and community, which was followed by the ACM Conference on Fairness, Accountability, and Transparency (ACM FAccT).} Transparency can benefit from the open publication of AI systems (including code, data, and models), though privacy, safety, and competitive considerations prevent this from being appropriate in all cases.\footnote{The necessity and sufficiency of transparency as an ideal for technical systems has also been critiqued in recent years, such as from Ananny and Crawford \cite{Ananny2018} and Kroll et. al. \cite{Kroll2017}.} Realizing transparency in AI development requires attention to institutional mechanisms and legal structures, particularly when scrutinizing developers' incentives and scope of activities. 

\textbf{Trust} and \textbf{trustworthiness}: these concepts have been extensively explored by researchers, though a consensus account across domains remains elusive. Substantial prior and ongoing work focuses on how these concepts manifest in AI development. This includes academic work \cite{Toreini2019},\footnote{See \cite{PAI2019} for a useful literature review.} government-associated efforts such as the European Union's High-Level Expert Group on AI \cite{High-LevelExpertGrouponAI2019}, and industry efforts \cite{GooglePAIR} \cite{Arnold2018}.  Our report focuses on a particular subset of what these concepts entail in the practice of AI development, namely the verifiability of claims about AI development, and more specifically, the verifiability of claims about safety, security, privacy, and fairness.

Several common frameworks for thinking about trust suggest a premium on the verifiability of claims even when they do not reference these terms explicitly. For example, Mayer et al.'s widely cited work \cite{Mayer1995} identifies benevolence, integrity, and ability as three pillars of trustworthiness. In the context of AI, a developer might make claims that suggest their pursuit of a benevolent goal (e.g., by adopting a set of ethical principles), but this needs to be accompanied by the skills and resources (ability) to achieve that goal as well as sufficient transparency and incentives to ensure consistent follow-through (integrity). 

Another prominent definition of trust which supports a focus on verifiable claims comes from Gambetta (paraphrased below):\footnote{This is a modification of Gambetta's definition \cite{gambetta1988trust}.}
\begin{displayquote}
"When we say we trust someone or that someone is trustworthy, we implicitly mean that we assess that the probability [they] will take actions that are beneficial (or at least not detrimental) is high enough for us to consider engaging in some form of cooperation with [them]."
\end{displayquote}

Here, too, the ability to scrutinize the claims and commitments made by an AI developer can provide calibration regarding the extent to which trust is appropriate in a given context.  

The verifiability of claims is also a key theme in the study of trust in the context of international relations and arms control \cite{coe_vaynman2019}. Ronald Reagan's famous "trust but verify" (a proverb taught to him by advisor on Russian affairs Suzanne Massie \cite{massie2016}) emphasized  the value of generating and assessing evidence of compliance with arms control agreements between the United States and Soviet Union. AI, verification, and arms control are discussed further in \hyperref[subsec:arms]{\textbf{Appendix IV}}.

This report is not intended to make novel contributions to the theory of trust or trustworthiness, but rather to explore verifiable claims as a building block of trustworthy AI development. When we use terms such as "earn trust" or "calibrate trust" in the report, these are meant to refer to cases in which evidence is provided to substantiate claims about an actor's behavior (including AI development).

\newpage

\subsection{The Nature and Importance of Verifiable Claims}\label{subsec:importance}

\textbf{Verifiable\footnote{While the report does discuss the technical area of formal verification at several points, the sense in which we use "verifiable" is distinct from how the term is used in that context. Unless otherwise specified by the use of the adjective "formal" or other context, this report uses the word verification in a looser sense. Formal verification seeks mathematical proof that a certain technical claim is true with certainty (subject to certain assumptions). In contrast, this report largely focuses on claims that are unlikely to be demonstrated with absolute certainty, but which can be shown likely or unlikely to be true, i.e. trustworthy, or untrustworthy through relevant arguments and evidence.} claims} are statements for which evidence and arguments can be brought to bear on the likelihood of those claims being true. Verifiable claims are sufficiently precise to be falsifiable, and the degree of attainable certainty in such claims will vary across contexts.

AI developers regularly make claims regarding the properties of AI systems they develop as well as their associated societal consequences. Claims related to AI development might include, e.g.:
\begin{itemize}
    \item We will adhere to the data usage protocols we have specified;
    \item The cloud services on which our AI systems run are secure;
    \item We will evaluate risks and benefits of publishing AI systems in partnership with appropriately qualified third parties;
    \item We will not create or sell AI systems that are intended to cause harm;
    \item We will assess and report any harmful societal impacts of AI systems that we build; and
    \item Broadly, we will act in a way that aligns with society's interests.
\end{itemize}

The verification of claims about AI development is difficult in part due to the inherent complexity and heterogeneity of AI and its supporting infrastructure. The highly dispersed ecosystem means there are many actors and sets of incentives to keep track of and coordinate. Further, the speed of development also means there is less time for claims to be carefully expressed, defended, and evaluated. And, perhaps most critically, claims about AI development are often too vague to be assessed with the limited information publicly made available. 

Notwithstanding these challenges, there are at least three distinct reasons why it is highly desirable for claims made about AI development to be verifiable. 

First, those potentially affected by AI development--as well as those seeking to represent those parties' interests via government or civil society--deserve to be able to scrutinize the claims made by AI developers in order to reduce risk of harm or foregone benefit.

Second, to the extent that claims become verifiable, various actors such as civil society, policymakers, and users can raise their standards for what constitutes responsible AI development. This, in turn, can improve societal outcomes associated with the field as a whole. 

Third, a lack of verifiable claims in AI development could foster or worsen a "race to the bottom" in AI development, whereby developers seek to gain a competitive edge even when this trades off against important societal values such as safety, security, privacy, or fairness \cite{Askell2019}. In both commercial (e.g., autonomous vehicles) and non-commercial (e.g., military) contexts, verifiable claims may be needed to foster cooperation rather than race-like behavior.

Without the ability to verify AI-related claims, the decision of how and whether to interact with AI systems must be made without information that could bear on the desirability of having that interaction. Given the large (and growing) stakes of AI development, such an information deficit is ethically untenable. An environment of largely unverifiable claims about AI could encourage extreme reactions to AI in particular situations (i.e., blind trust or blind rejection), resulting in both over-use and under-use of AI.  The world instead needs trust in AI development to be well-calibrated, i.e. it should be the case that confidence in certain claims or actors is proportional to the available evidence. The benefits and risks of AI are many, and need to be appraised with context and nuance.

In the past decade, claim-oriented approaches have been developed in order to structure arguments about the safety of engineered systems \cite{Bloomfield2019},  and we draw inspiration from such approaches in this report. One result of such work is the introduction and standardization of assurance cases in numerous domains. An assurance case is a documented body of evidence that provides a convincing and valid argument regarding a top-level claim (such as the safety of a nuclear power plant), and presents a structured justification in support of that claim to decide the status of it. Assurance cases are often required as part of a regulatory process (e.g., a certificate of safety being granted only when the regulator is satisfied by the argument presented in a safety case).\footnote{Assurance cases are primarily concerned with demonstrating the validity (or otherwise) of the resulting argument and have two main roles: logical reasoning and communication. Cases are usually integrated within a regulatory process that provides for independent challenge and review. There can be a number of stakeholders including public officials, developers, certifiers, regulators. Communication is thus essential to create a shared understanding between the different stakeholders, build confidence and consensus}

This work matured into the widely-used Claims, Arguments, and Evidence (CAE) framework.\footnote{See, e.g., this discussion of CAE in the nuclear safety context \cite{OfficeforNuclearRegulationONR2019} and Uber's use of GSN \cite{Uber}} CAE is often the framework of choice in aviation, nuclear, and defense industries worldwide to reason about safety, security, reliability and dependability, and recent work has begun applying CAE to the safety analysis of AI systems.\footnote{See Zhao et al. \cite{Zhao2020}.} 

The CAE framework consists of three key elements. \textbf{Claims} are assertions put forward for general acceptance. They're typically statements about a property of the system or some subsystem. Claims asserted as true without justification are assumptions, and claims supporting an argument are subclaims. \textbf{Arguments} link evidence to a claim, which can be deterministic, probabilistic, or qualitative.\footnote{Arguments are presented in a form of defeasible reasoning of top-level claims, supported by the available evidence; and driven by practical concerns of achieving the required goal in the best possible way considering the existing uncertainties, point of views, concerns and perspectives of different stakeholders. Such argumentation is expected to be multidisciplinary, and cover a wide range of mechanisms, which we aim to address. To support CAE, a graphical notation can be used to describe the interrelationship of claims, arguments, and evidence. Claim justifications can be constructed using argument blocks--concretion, substitution, decomposition, calculation, and evidence incorporation—as well as narrative analyses that describe the claims, arguments, and evidence in detail.} They consist of "statements indicating the general ways of arguing being applied in a particular case and implicitly relied on and whose trustworthiness is well established" \cite{toulmin1979}, together with validation of any scientific laws used. In an engineering context, arguments should be explicit. \textbf{Evidence} serves as the basis for justification of a claim. Sources of evidence can include the design, the development process, prior experience, testing, or formal analysis.

For a sufficiently complex AI system or development process, a wide variety of mechanisms will likely need to be brought to bear in order to adequately substantiate a high-level claim such as "this system was developed in accordance with our organization's ethical principles and relevant laws."

\newpage

\subsection{AI, Verification, and Arms Control}\label{subsec:arms}

At an international level, arms control is a possible approach to addressing some of the risks of AI development in a military context. Arms control involves similar issues to those discussed earlier (namely, the need for credible commitments and close attention to transparency and incentives) in non-military contexts. In this subsection, we provide an overview of the relationship between verifiable claims and arms control applied to AI. 

Arms control is a special case of regulation, in which nation-states cooperate to self-regulate weapons technologies under particularly challenging conditions \cite{Brennan1961} \cite{Brodie1962} \cite{burns1993encyclopedia} \cite{Burns2013} \cite{Goldblat1982}. Unlike in domestic regulation, there is no external actor to force compliance if states violate the terms of an arms control agreement. Instead, states generally rely on reciprocity to enforce arms control agreements. If states violate an agreement, they can often expect others to follow suit and develop the weapon themselves. 

Formal agreements such as treaties act as coordination mechanisms for states to reach agreement, but do not directly enforce compliance. Some treaties include verification regimes to help increase visibility among states as to whether or not others are complying with an agreement, as a means of facilitating trust \cite{Caughley2016} \cite{Gallagher2003}. But it is on states themselves to take action if others are found violating an agreement, whether through sanctions, reciprocal weapons development, military action, or other tools of statecraft.

Arms control is inherently challenging not only because there is no third-party enforcement mechanism, but because states may be incentivized to violate agreements if they believe that doing so may give them an edge against competitors \cite{Herz1952} \cite{Jervis1978} \cite{Jervis1993}. This tension is exacerbated if it is challenging to verify other states' behavior. States may assume others are cheating and developing a prohibited technology in secret, incentivizing them to do so as well or risk falling behind a competitor. Arms control agreements can also be more challenging to hold together if a technology is more widely accessible to a larger number of actors and if defection by one actor generates incentives for others to defect. There are many cases in history in which nation-states genuinely desired mutual restraint for certain weapons, such as turn of the century rules regulating submarines, air-delivered weapons, and poison gas, but states were unable to achieve effective cooperation in wartime for a variety of reasons.

Despite these hurdles, there have been successful examples of arms control for a number of weapons, including: chemical \cite{ChemicalWeaponsConvention} \cite{Brown1968} and biological \cite{barner-barry_1990} weapons; land mines \cite{Rutherford2000} \cite{UnitedNations1997}; cluster munitions \cite{UnitedNations2008}; blinding lasers \cite{Roff2016}; exploding bullets; limits on the proliferation, quantity, and deployment of nuclear weapons \cite{AtomicHeritageFoundation2017}; anti-ballistic missile systems \cite{ABM_Treaty}; weapons of mass destruction in space; and weapons on the Moon or in Antarctica. There are also examples of mutual restraint with some weapons despite the lack of formal agreements, including neutron bombs, kinetic (debris-causing) anti-satellite weapons, and certain forms of bayonets. 

Even these successes highlight the limitations of arms control, however. Some treaties have collapsed over time as more nations gained access to the underlying technology and did not abide by the prohibition. And even the most successful prohibitions, such as those on chemical and biological weapons, have failed to rein in rogue regimes or terrorists. Despite the widespread global condemnation of chemical weapons, Bashar al Assad has used them in Syria to murder civilians, with minimal consequences from the international community.

In general, arms control is more likely to succeed when:\footnote{See, e.g., Crootof \cite{Crootof2015}, Watts \cite{Watts2015}, and Scharre \cite{Scharre2019}.} (1) there are clear lines between which weapons are prohibited and which are permitted; (2) the perceived horribleness of a weapon outweighs its military value; (3) states have the ability, either through formal verification regimes or other mechanisms, to ensure that others are complying with the regulation; and (4) fewer states are needed for an agreement to work. Regulation can occur at multiple points of technology development, limiting or prohibiting access to the underlying technology, weapons development, production, and/or use. Note also that while some of the variables above are exogenous from the perspective of the AI community, others are potentially amenable to influence (e.g., research could potentially improve the distinguishability of offensive and defensive uses, or improve the general traceability and interpretability of AI systems).

AI will likely be a transformative technology in warfare \cite{payne2018}. Anticipating this transformation, many in the scientific community have called for restrictions or bans on military AI applications. Because AI is a general-purpose enabling technology with many applications, broad bans on AI overall are unlikely to succeed. However, prohibitions on specific military applications of AI could succeed, provided states could agree to such limits (requiring that the terms be compatible with the incentives of each party) and that appropriate means of verifying compliance are developed and implemented.

AI technology has certain attributes that may make successful restraint challenging, however. These include its widespread availability, dual use or "omni-use" nature \cite{Brundage2018}, the difficulty in drawing clear lines between acceptable and unacceptable AI applications, and the challenges of verifying compliance, which are at least as difficult as those found in non-military contexts and perhaps more challenging given the more adversarial context.

One special case worth highlighting is the development of lethal autonomous weapon systems (LAWS). An international LAWS ban has been the subject of discussion at the UN Convention on Certain Conventional Weapons (CCW) since 2014. There are many arguments made in support of restrictions on LAWS. Three relevant arguments are: (1) their use is immoral because AI systems will not in the foreseeable future understand the moral, psychological, and social context at the time of killing a person (unlike a human, who could decide to not press the trigger) \cite{InternationalCommitteoftheRedCross2014} \cite{Bhuta2016}; (2) the state of the technology today would preclude their use under international law in anything but isolated cases, such as undersea where civilians are not present; and (3) they might proliferate easily, enabling misuse \cite{Brundage2018} \cite{Allen2017}. Those skeptical of a ban on lethal autonomous weapon systems often reference mutual distrust as a reason for development: "if we don't develop them, others will, putting us at a strategic disadvantage" is a refrain echoed by several great powers.

Avoiding such an impasse requires grappling with the issue of trust head-on, and closely attending to the complexities of AI development in practice. Similar to how AI ethics principles need to be supplemented with mechanisms that demonstrate the implementation of such principles, trust in military-relevant AI systems must be supported by mechanisms based on a rigorous analysis of the dynamics of military AI development. Lethal autonomous weapons are currently the focus of much related discussion, though the use of AI in cyberwarfare and nuclear command and control have also been highlighted as challenging areas in recent years. Some early work in the direction of coordination on AI among great powers \cite{Imbrie2019} has called attention to the need for early dialogue on AI safety and security. Other work has fleshed out the notion of meaningful human control as a cornerstone of lethal autonomous weapon system governance \cite{Moyes2016} \cite{Roff2016} \cite{Roff2017}.

The AI community and advocates in other disciplines have played a key role in bringing this issue to the attention of the international community \cite{Belfield2020} \cite{carpenter2014} \cite{bahcecik2019} \cite{verbruggen2019} \cite{Vignard2018} \cite{Frederick2019} \cite{FutureofLifeInstitute2015}. Similar efforts by expert communities have improved prospects for arms control in prior contexts such as nuclear weapons \cite{Adler1992} \cite{Haas1992}. There remains more to be done to raise the profile of the issue among policymakers, and to identify appropriate steps that individuals and organizations in the AI community can take to forestall the development of potentially harmful systems. 

AI researchers could contribute technical expertise that helps identify potential governance mechanisms in this context. For example, AI researchers, working with arms control experts in other disciplines, could scrutinize proposals such as defensively-oriented AI weapons systems that could target lethal autonomous weapons (but not humans) and help think through different means of limiting proliferation and ensuring human accountability. The AI community's distributed expertise in the process of AI development and the feasibility of different technical scenarios could thus be brought to bear to limit AI "arms racing" and prevent a race to the bottom with respect to the safety, security, and human-accountability of deployed military AI systems \cite{Scharre2019}. The feasibility of verifying, interpreting, and testing potential AI systems designed for various purposes, as well as the feasibility of using different points of control for governance of the supply chain (e.g., the computing and non-computing hardware associated with autonomous weapons vs. the underlying software), are all issues to which AI expertise is relevant.

Of the various inputs into AI development (including hardware, software, data, and human effort), it's worth noting that hardware is uniquely governable, at least in principle. Computing chips, no matter how fast, can perform only a finite and known number of operations per second, and each one has to be produced using physical materials that are countable, trackable, and inspectable.\footnote{We emphasize that this discussion is exploratory in nature, and that there would be major practical challenges involved in acting on these high-level ideas. Our goal in highlighting the unique affordances of hardware is to foster creative thinking about these issues rather than to suggest that there is a readily available solution to the weaponization of AI.} Similarly, physical robots rely on supply chains and materials that are in principle trackable. Computing power and hardware platforms for robotics are thus potentially amenable to some governance tools used in other domains that revolve around tracking of physical goods (e.g., export controls and on-site inspections). 

While it is unclear what hardware-based verification efforts might look like in the context of AI-related arms control, and how feasible they would be, one might imagine, e.g., a bottleneck in the manufacturing process for lethal autonomous weapons. In contrast to such a bottleneck, AI-related insights, data, code, and models can be reproduced and distributed at negligible marginal cost, making it inherently difficult to control their spread or to use them as a metric for gauging the capabilities of an organization with respect to developing lethal decision-making systems.\footnote{Another potential approach would be to impose constraints on the physical characteristics of AI-enabled military systems, such as their range, payload, endurance, or other non-AI related physical attributes.} Given such considerations, it is incumbent upon stakeholders to consider ways in which the distinctive properties of hardware might be leveraged in service of verifying any future arms control agreements.

\newpage

\subsection{Cooperation and Antitrust Laws}\label{subsec:laws}

Collaborations between competing AI labs, even for beneficial purposes such as enabling verifiable claims, can raise antitrust issues. Antitrust law is also known as "competition law" or "anti-monopoly law" outside the US. This section primarily addresses US antitrust law, but given the international nature of AI development and markets, attention to the international legal implications of industry collaborations is warranted. 

US antitrust law seeks to prevent "unreasonable" restraints on trade \cite{StandardOilUS1911}. Unreasonableness, in turn, is tested by economic analysis \cite{Kovacic2000}--specifically, a "consumer welfare" test \cite{Orbach2011}. Although recent academic \cite{Khan2017} and popular \cite{Herndon2019} proposals challenge the wisdom and usefulness of this test, consumer welfare remains the guiding principle for antitrust courts \cite{Orbach2011}.

Antitrust law generally condemns per se particularly harmful restraints on trade,\footnote{Practices are condemned per se "[o]nce experience with [that] particular kind of restraint enables the Court to predict with confidence that [antitrust analysis] will condemn it..."\cite{ArizonaMaricopa1982}} such as direct restraints on price, output levels, competitive bidding, and market allocation \cite{PracticalLawAntitrust}. Other practices are analyzed by the Rule of Reason, "according to which the finder of fact must decide whether the questioned practice imposes an unreasonable restraint on competition, taking into account a variety of factors, including specific information about the relevant business, its condition before and after the restraint was imposed, and the restraint's history, nature, and [net] effect." \cite{StateOilKhan1997} Importantly, courts in the past have consistently \textbf{rejected} safety-based (and other public policy-based) defenses of anticompetitive behavior \cite{FTCdentists1986}.

In a leading case on point, \textit{National Society Professional Engineers v. United States}, the US Supreme Court reasoned that by adopting antitrust laws, Congress had made a "basic policy" decision to protect competition \cite{NatSocUS1978}. The Court therefore concluded that the defendant's argument that price competition between engineers was unsafe "[wa]s nothing less than a frontal assault on the basic policy of the [antitrust laws].\footnote{\textit{See id}. at 695.}"  "In sum," the Court held, "the Rule of Reason does not support a defense based on the assumption that competition itself is unreasonable."\footnote{\textit{Id} at 696.}

None of this implies, however, that collaborations between competitors are always anticompetitive and therefore violative of antitrust laws. American antitrust authorities have acknowledged that collaborations between competitors can have important procompetitive benefits, such as enabling new products to be developed, sharing useful know-how, and capitalizing on economies of scale and scope \cite{FederalTradeCommision2000}. These benefits need to be balanced against possible harms from collaboration such as reduced competition on pricing or output levels, reducing the pace of progress, or increasing the uniformity of outputs.\footnote{\textit{See generally id.}}

If the right antitrust governance procedures are in place, joint activities between competitive AI labs can both enhance consumer welfare and enhance intra-industry trust. Nevertheless, it is important to not allow the goal of supporting verifiable claims to provide cover for practices that would harm consumer welfare and therefore erode trust between society and AI labs collectively.

\newpage

\subsection{Supplemental Mechanism Analysis}\label{subsec:analysis}

\subsubsection{Formal Verification}

Formal verification techniques for ML-based AI systems are still in their infancy. Challenges include:
\begin{itemize}
    \item Generating formal claims and corresponding proofs regarding the behavior of ML models, given that their output behavior may not always be clear or expected relative to the inputs (e.g., an ML model will not necessarily display the same behavior in the field that it exhibited under a testing environment). As a consequence, traditional formal properties must be reconceived and redeveloped for ML models;
    \item The difficulty of correctly modeling certain ML systems as mathematical objects, especially if their building blocks cannot be formalised within mathematical domains utilized by existing verification techniques; and
    \item The size of real-world ML models, which are usually larger than existing verification techniques can work with.
\end{itemize}

Some preliminary research \cite{Bastani2017} has attempted to find ways of specifying types of ML robustness that would be amenable to formal verification: for example, pointwise robustness. Pointwise robustness is a property that states that an ML model is robust against some model of adversarial attacks and perturbations at a given point \cite{Goodfellow2015}. However, researchers \cite{Bloomfield2019} have observed that the maturity and applicability of both the specification and corresponding techniques fall short of justifying functionality, dependability, and security claims. In general, most system dependability properties have gone unspecified,\footnote{For example:  functionality, performance, reliability, availability, security, etc.} and these methodologies have not accounted for specifications that are more unique to ML-based systems.

Other efforts \cite{Katz2017} \cite{Pulina2010} \cite{Liu2019} aim to verify more traditional specifications regarding ML algorithms. Some of these techniques require functional specifications, written as constraints, to be fed into specialized solvers which then attempt to verify that they hold on a constraint model of the ML system. The generalization of these techniques to deep learning is challenging because they require well-defined, mathematically specifiable properties as input which are not unique to ML algorithms (given that such properties do not easily lend themselves to such specifications). These techniques are only applicable to well-specified deterministic or tractable systems that can be implemented using traditional methods (e.g., the C programming language) or via ML models. As a consequence, these techniques cannot be straightforwardly applied to arbitrary contexts, and domain-specific effort is currently required even to specify properties of interest, let alone verify them.

Indeed, there is much progress to be made with regard to the verification of deep neural networks, but formal verification can still be effectively utilised to reinforce non-ML software employed to construct the ML model itself. For example, researchers have demonstrated a methodology in which developers can use an interactive proof assistant to implement their ML system and prove formal theorems that their implementation is free of errors \cite{Selsam2017}. Others have shown that overflow and underflow errors within supporting software can propagate and affect the functionality of an ML model \cite{Hutchison2018}. Additionally, researchers have identified a number of different run-time errors using a traditional formal methods-based static-analysis tool to analyze YOLO, a commonly used open source ML vision software \cite{Bloomfield2019}. Issues identified include:
\begin{itemize}
    \item A number of memory leaks, such as files opened and not closed, and temporarily allocated data not freed, leading to unpredictable behavior, crashes, and corrupted data;
    \item A large number of calls to free where the validity of the returned data is not checked. This could lead to incorrect (but potentially plausible) weights being loaded to the network;
    \item Potential "divide by zeros" in the training code. This could lead to crashes during online training, if the system were to be used in such a way; and
    \item Potential floating-point "divide by zeros," some of which were located in the network cost calculation function. As noted above, this could be an issue during online training.
\end{itemize}

We note that many of the above errors are only applicable to languages such as C and C++ (i.e., statically typed languages), and not Python, a language widely used in the implementation of numerous ML libraries and frameworks. As a dynamically typed language, Python brings about a different set of program errors not typically exhibited by statically typed languages (e.g., type errors). Unfortunately, formal verification techniques for the analysis of Python code are inherently limited, with linters and type checkers being the main available source of static analysis tools.

Though the Python situation differs from that encountered with C and C++, there are many ways that potential faults arising from Python could affect the functionality of an ML model. This is a large gap within the formal verification field that needs to be addressed immediately, given the deployment of safety-critical AI systems, such as autonomous vehicles, utilizing Python. Previous research efforts\footnote{See Python semantics: (\url{https://github.com/kframework/python-semantics}).} \cite{Politz2013} have attempted to formalise a subset of Python that would be amenable to verification; however, it has been notoriously difficult to formalise and verify \cite{Gardner2012} dynamically typed languages. Although optional static type hinting is now available for Python,\footnote{See official Python documentation (\url{https://docs.python.org/3/library/typing.html}) and MyPy (\url{https://github.com/python/mypy}).} "the Python runtime does not enforce function and variable type annotations. [Hints] can be used by third party tools such as type checkers, IDEs, linters, etc." Furthermore, it is unlikely that the ML community will constrain themselves to subsets of Python which are statically-typed.\footnote{See the discussion following the feature request in the TensorFlow codebase (\url{https://github.com/tensorflow/tensorflow/issues/12345}).}

Formal verification techniques have been widely deployed for traditional safety-critical systems (as required by IEC 61508) for several decades, and have more recently been adopted by some tech companies for specific applications.\footnote{See Infer, an open source static analyzer (\url{https://fbinfer.com/}).} However, the rapid introduction of machine learning in these environments has posed a great challenge from both a regulatory and system assurance point of view. The lack of applicable formal verification techniques for AI systems stifles the assurance avenues required to build trust (i.e., regulations, verification, and validation), curbing the potential innovation and benefits to be gained from their deployment. The following open research problems must thus be addressed to allow formal verification to contribute to trust in AI development:
\begin{itemize}
    \item Creation of specifications unique to AI, with corresponding mathematical frameworks, to contribute to assurance of AI systems;
    \item Creation of novel formal verification techniques which can address the newly defined specifications mentioned above; and
    \item Collaboration between ML and verification researchers resulting in deep learning systems that are more amenable to verification \cite{Kuper2018}.
\end{itemize}

\newpage

\subsubsection{Verifiable Data Policies in Distributed Computing Systems}\label{subsec:distributed}

Current IT systems do not provide a mechanism to enforce a data policy (e.g., sharing restrictions, anonymity restrictions) on data that is shared with another party - individuals and organizations are required to trust that the data will be used according to their preferences. Google's Project Oak\footnote{More detail on Oak's technical aspects, including instructions for how to write programs targeting the current iteration of Oak, can be found on the Oak GitHub repo: \url{https://github.com/project-oak}} aims to address this gap, by providing a reference implementation of open source infrastructure for the verifiably secure storage, processing, and exchange of any type of data. 

With Oak, data is collected and used as it is today, but it is also accompanied by enforceable policies that define appropriate access, collection, and use. Data is stored in encrypted enclaves and remote attestation between enclaves ensures that only appropriate code ever gets direct access to the secured data  (i.e. within the limits of what can be verified and as defined by a configurable policy), and processing of the data creates a verifiable record. In the long term, Google's objective is to provide formal proofs such that core properties of the system can be verified down to the hardware. Platforms that implement this infrastructure could then form the bedrock of all sorts of other services from messaging, machine learning, and identity management to operating system hosting, making meaningful control technically feasible in a way that it is not today.

Oak uses enclaves and formal verification. Taken together, it is possible to verify that data is only processed in a way that complies with a configurable policy that goes with it. In short, data lives in enclaves and moves from one enclave to another only when the sending enclave is able to convince itself that the receiving enclave will obey the policy that goes with the data and will itself perform the same verification step before sending the data (or data derived from it) on to other enclaves. Movement outside enclaves is only permitted when encrypted with keys available only to the enclave, or as allowed by policy (for example, to show the data to specific people or when adequately anonymized, again specified by policy).

Oak combines formal verification and remote attestation with binary transparency. Oak is being developed entirely as an open source project - this is deliberate and necessary. Because Oak is open source, even in the absence of formal proofs, any independent third party (whether an individual researcher, regulatory body, or consumer advocacy group) can examine Oak's source code and confirm that the implementation matches the expected behavior. With the correctness of Oak confirmed insofar as possible, a given Oak virtual machine needs to be able to attest that it is running the "correct" binary of Oak. This attestation makes it possible for the client (or sending enclave) to assure itself that the requesting enclave will follow any policies on the data, because it knows that the Oak policy enforcement system is running on that enclave and is "truly" Oak - that is: matches the binary of Oak known from the open source repository.

Usability and independent auditability are crucial to an Oak system's utility. Four types of actors are expected to interact with Oak:
\begin{itemize}
    \item End users: people who will use Oak apps;
    \item Application developers: people who will build Oak apps;
    \item Policy authors: people who will define and manage the policies that accompany the data in an Oak app;
    \item Verifiers: people who will add credibility to an Oak app by verifying that the policy is upheld.
\end{itemize}

A user-centric design perspective points to many questions such as: What do people need to know about an Oak app when making a decision about whether to use it? How will people understand the effective policy an Oak app is bound by? How will people's preferences be captured? How will we help people find verifiers to delegate trust decisions to? If Oak apps and data policies change, or a verifier's assessment of a policy changes (which we expect can and will happen), how is this change communicated to the people who need to know? 

As important as it is to ensure end-users avoid critical mistakes and can understand the impact of Oak, it is even more important to ensure developers are able to avoid critical mistakes when using Oak. This requires deliberate design of how app makers will build, deploy, and debug.

An Oak node without a correct and useful policy is useless. Oak does not provide privacy by default, it does so only if the policies specified are privacy-preserving. Thus, the user experience of specifying and maintaining the policy that accompanies data is crucial to the successful use of Oak. Policy authors will begin with a set of policy goals that will be refined into a natural language representation of a set of rules and will likely be translated into a set of rules that can be enforced by an Oak node. The policy language for those rules will need to be determined based on the types of protection that is relevant to use cases and the rules that can be verifiably enforced. In some cases, there will be desirable attestations that cannot be formally verified (e.g., non-commercial use). Depending on the context, policies may be generated and managed by a group of people, sometimes the developer and sometimes a cross-functional team from within a company. 

\newpage

\subsubsection{Interpretability}\label{subsec:interpretability}

\textbf{What has interpretability research focused on?}

Interpretability research includes work in areas such as explaining a specific prediction \cite{Koh2017} \cite{Khanna2018} \cite{Sharchilev2018} \cite{Yeh2018}, explaining global model behavior \cite{Zhang2019} \cite{Bastani2017} \cite{Lakkaraju2016} \cite{Tan2018} \cite{Zhang2017a}, building more interpretable models \cite{Brendel2019} \cite{Zhang2017} \cite{Lakkaraju2016} \cite{Chen2016} \cite{Higgins2017} \cite{Verma2018} \cite{Ross2017}, interactive visualization tools for human exploration \cite{Amershi2015} \cite{Wu2019} \cite{Berner2019} \cite{Vig2019} \cite{Bau2019a}, and analyzing functional sub-components of neural networks to understand what they are doing \cite{Carter2019}. These areas of work are characterized separately below, although several overlap and interact with one another.

\textbf{Explaining a specific prediction}. Many techniques seek to explain a model's prediction on some given input. For example, one might ask which part of the input--for image models, this might take the form of a heatmap over input pixels \cite{Springenberg2014} \cite{Zeiler2014} \cite{Fong2017} \cite{Fong2019} \cite{Dabkowski2017} \cite{Petsiuk2018} \cite{Simonyan2013}--or which training examples \cite{Koh2017} \cite{Khanna2018} \cite{Sharchilev2018} \cite{Yeh2018} were responsible for the model's prediction. By examining the model's reasoning on different input instances through these attribution methods, we can better decide whether or not to trust the model: if a model, say, predicts that an image contains a wolf because of the snowy image background and not because of any actual wolf features, then we can extrapolate that it is likely to misclassify future images with snowy backgrounds \cite{Ribeiro2016a}.

\textbf{Explaining global model behavior}. Instead of explaining individual predictions, other techniques aim to construct human-understandable representations of a model's global behavior, so that users can more directly interrogate what a model might do on different inputs rather than having to extrapolate from explanations of previous predictions. Examples include approximating a complex model with a simpler, more interpretable model (like a shallow decision tree) \cite{Zhang2019} \cite{Bastani2017} \cite{Lakkaraju2017a} \cite{Tan2018} \cite{Zhang2017a}; or characterizing the role and contribution of internal components of a model (e.g., feature visualization or development of geometric invariance) \cite{Zeiler2014} \cite{Simonyan2013} \cite{Mahendran2016} \cite{Nguyen2016} \cite{Bau2017} \cite{Olah2017} \cite{Ulyanov2018} \cite{Olah2018} \cite{Morcos2018} \cite{Lenc2014}.

\subsubsection*{\textit{Current directions in interpretability research}}

\textbf{Building more interpretable models}. A separate but complementary line of work seeks to build models that are constrained to be interpretable by design (as opposed to training a complex, hard-to-interpret model and then attempting to analyze it post-hoc with one of the above techniques) \cite{Brendel2019} \cite{Zhang2017} \cite{Lakkaraju2016} \cite{Chen2016} \cite{Higgins2017} \cite{Verma2018} \cite{Ross2017}.

\textbf{Interactive visualization tools for human exploration}. A related research theme is the development of tools that allow humans to interact with, modify, and explore an ML system (e.g., dashboards to visualize model predictions and errors \cite{Amershi2015} \cite{Wu2019}; explanations of model representations \cite{Vig2019} \cite{Bau2019a} \cite{VanDerMaaten2008} \cite{Cai2019}; or directly interacting with an AI agent \cite{Berner2019}.\footnote{Also see Google Quickdraw (\url{https://quickdraw.withgoogle.com/}) and Bach doodle (\url{https://www.google.com/doodles/celebrating-johann-sebastian-bach}).}).

\textbf{Software and tools for practitioners}. Most interpretability tools that have been developed are best used by ML researchers for interpretability research. A few software packages that are less research-oriented allow novice users to better understand a dataset's distribution and inspect a model interactively;\footnote{See What-if tool for ML model exploration (\url{https://pair-code.github.io/what-if-tool/}) and Facets for data exploration (\url{https://pair-code.github.io/facets/}).} these packages primarily fall under the category of "interactive visualization tools." Moreover, most open-sourced code from interpretability research primarily focuses on the method being introduced and rarely include standardized benchmarks and comparisons with related work, with some exceptions.\footnote{See saliency repository (/url{https://github.com/PAIR-code/saliency}), InterpretML (\url{https://github.com/interpretml/interpret}), and TreeInterpreter (\url{https://github.com/andosa/treeinterpreter}).} We hope to see more software packages that empower novice users to use interpretability techniques effectively as well as aid researchers by providing standardized benchmarks for comparing methods. Lastly, much work is focused on interpretability at a particular scale (i.e., individual examples vs. dataset distribution); we desire more work at connecting interpretability work along different axes and scales \cite{Olah2018}.

\newpage
{\Large List of Recommendations for Reference}

\hyperref[sec:institutional]{\textbf{Institutional Mechanisms and Recommendations}}
\begin{enumerate}
    \item A coalition of stakeholders should create a task force to research options for conducting and funding \hyperref[subsec:audit]{\textbf{third party auditing}} of AI systems.
    \item Organizations developing AI should run \hyperref[subsec:redteam]{\textbf{red teaming exercises}} to explore risks associated with systems they develop, and should share best practices and tools for doing so.
    \item AI developers should pilot \hyperref[subsec:bounties]{\textbf{bias and safety bounties}} for AI systems to strengthen incentives and processes for broad-based scrutiny of AI systems.
    \item AI developers should share more information about \hyperref[subsec:incidents]{\textbf{AI incidents}}, including through collaborative channels.
\end{enumerate}
\hyperref[sec:software]{\textbf{Software Mechanisms and Recommendations}}
\begin{enumerate}[resume]
    \item Standards setting bodies should work with academia and industry to develop \hyperref[subsec:trails]{\textbf{audit trail}} requirements for safety-critical applications of AI systems.
    \item Organizations developing AI and funding bodies should support research into the \hyperref[subsec:interpretabilitysoft]{\textbf{interpretability}} of AI systems, with a focus on supporting risk assessment and auditing.
    \item AI developers should develop, share, and use suites of tools for \hyperref[subsec:ppml]{\textbf{privacy-preserving machine learning}} that include measures of performance against common standards.
\end{enumerate}
\hyperref[sec:hardware]{\textbf{Hardware Mechanisms and Recommendations}}
\begin{enumerate}[resume]
    \item Industry and academia should work together to develop \hyperref[subsec:securehardware]{\textbf{hardware security features}} for AI accelerators or otherwise establish best practices for the use of secure hardware (including secure enclaves on commodity hardware) in machine learning contexts.
    \item One or more AI labs should estimate the computing power involved in a single project in great detail (\hyperref[subsec:measurement]{\textbf{high-precision compute measurement}}), and report on the potential for wider adoption of such methods.
    \item Government funding bodies should substantially increase \hyperref[subsec:support]{\textbf{funding of computing power resources}} for researchers in academia, in order to improve the ability of those researchers to verify claims made by industry.
\end{enumerate}

\end{document}